\documentclass[aps, pra, preprint, groupedaddress, amsfonts,
               amsmath, amssymb, nofootinbib]{revtex4-1}
\usepackage{microtype}
\usepackage{graphicx}
\usepackage{epstopdf}
\usepackage[utf8]{inputenc}
\usepackage[T1]{fontenc}
\usepackage[usenames,dvipsnames]{xcolor}
\usepackage{hyperref}

\newcommand{\bra}[1]{\langle #1|}
\newcommand{\ket}[1]{|#1\rangle}
\newcommand{\braket}[2]{\langle #1|#2\rangle}

\usepackage{color}

\begin{document}
\title{Independent-atom-model coupled-channel calculations 
strengthen the case for interatomic Coulomb decay as
a subdominant reaction channel in slow O$^{3+}$-Ne$_2$ collisions}

\author{Dyuman Bhattacharya}
\affiliation{Department of Physics and Astronomy, York University, Toronto, Ontario M3J 1P3, Canada} 

\author{Tom Kirchner}  
\email[]{tomk@yorku.ca}
\affiliation{Department of Physics and Astronomy, York University, Toronto, Ontario M3J 1P3, Canada}
\date{\today}
\begin{abstract}
We report on electron removal 
calculations for 2.81 keV/amu Li$^{3+}$ and O$^{3+}$ ion collisions with neon dimers.
The target is described as two independent neon atoms fixed at 
the dimer's equilibrium bond length, whose electrons are subjected to the time-dependent
bare and screened Coulomb potentials of the classically moving
Li$^{3+}$ and O$^{3+}$ projectile ions, respectively. Three mutually perpendicular
orientations of the dimer with respect to the rectilinear projectile trajectories
are considered and collision events for the two ion-atom subsystems are combined
in an impact parameter by impact parameter fashion and are orientation-averaged
to calculate probabilities
and cross sections for the ion-dimer system. 
The coupled-channel two-center basis generator method is used to solve
the ion-atom collision problems. 
We concentrate the ion-dimer analysis on one-electron and two-electron removal processes 
which can be associated with
interatomic Coulomb decay, Coulomb explosion, and radiative charge transfer.
We find that the calculated relative yields are in fair agreement with recent
experimental data for O$^{3+}$-Ne$_2$ collisions 
if we represent the projectile by a screened Coulomb potential, but disagree markedly
for a bare Coulomb potential, i.e., for Li$^{3+}$ impact.
In particular, our calculations suggest that interatomic Coulomb decay is a significant
reaction channel in the former case only, since capture of a Ne($2s$) electron to form
hydrogenlike Li$^{2+}$ is unlikely.
\end{abstract}
%
%

\maketitle
\section{Introduction}
\label{intro}
Rare-gas dimers are much studied objects of the microworld with fascinating
structural and dynamical properties.
Their (van der Waals) bonds are weak and their internuclear
distances large so that
the two atoms appear to be (quasi-) independent.
However, it has been demonstrated that charge and energy transfer
between the two sites are possible and do happen after excitation
by photon or charged-particle impact. 
Perhaps the most celebrated example of such a process is interatomic Coulomb decay
(ICD), which 
is initiated by the removal of an inner-valence electron
from one atom by the impinging particle or radiation.
ICD then involves the transfer of the
excitation energy to the other atom, its release in the 
form of (low-energy) outer-shell electron emission, and the fragmentation 
of the system of two singly-charged ground-state ions 
produced in this way.

ICD was predicted in 1997 based on ab-initio calculations~\cite{Cederbaum97}. 
The first experimental evidence
was reported in a study of
photoexcited neon clusters in 2003~\cite{Marburger03} and was 
unequivocally confirmed
for neon dimers one year later~\cite{Jahnke04}. 
Since then, a large number of theoretical and experimental studies have
provided further data and insight (see, e.g., Ref.~\cite{Jahnke15} and references therein).
ICD is now considered to be a ubiquitous process in a variety of systems,
and the associated low-energy electron emission is deemed to play an important role in
the radiation damage of biological tissue (see, e.g., Ref.~\cite{Ren18} and references therein).

ICD in neon dimers subjected to slow multiply-charged ion impact
was reported in Ref.~\cite{Iskandar15}. 
More specifically, kinetic energy release (KER) spectra for the
Ne$^+(2p^{-1})$ + Ne$^+(2p^{-1})$ fragmentation channel (we use the same 
notation as the authors of Ref.~\cite{Iskandar15} to indicate 
vacancies in a given atomic subshell) were recorded in
coincidence with the final projectile charge state, and peaks in those
spectra were associated with three different processes based on an
analysis involving some of the potential energy curves of the dimer system.
ICD resulting from the primary removal of one Ne($2s$) electron was
one of these processes. The other two were radiative charge transfer (RCT)
and Coulomb explosion (CE). The latter corresponds to the direct production
of Ne$^+(2p^{-1})$ + Ne$^+(2p^{-1})$ in the collision by electron capture
of one $2p$ electron from each atom, while the former is the result of a two-electron
capture process from one atom, producing a transient state which relaxes radiatively
to the same Ne$^+(2p^{-1})$ + Ne$^+(2p^{-1})$ channel as CE and ICD, but
involves higher KER values.
Relative yields for these processes were determined for three different projectile
species: O$^{3+}$, Ar$^{9+}$, and Xe$^{20+}$ ions.
While RCT and CE were found to contribute for all three projectiles, the
characteristic ICD peak was only 
present for O$^{3+}$ impact, contributing 20\% to the total yield.

These findings were supported by classical over-the-barrier model (COBM)
calculations published along with the data. The calculations were based on
an independent-atom-model (IAM) description of the ion-dimer collision problem using
bare Coulomb potentials for the projectiles. For the O$^{3+}$-Ne$_2$ system
they resulted in at most qualitative agreement with the measurements; in particular
the ICD channel appeared to be too weak (contributing just 8.2\% to the total yield), 
while the CE yield was found to be significantly stronger than in the experiment. 
Given that the O$^{3+}$-Ne$_2$ system was the only one that showed evidence for ICD, 
an independent
calculation based on a higher-level theory is desirable. This is the motivation
for the present work.

Our calculations are also based on the IAM, but the ion-atom collisions are
described in a semiclassical coupled-channel framework using the
two-center basis generator method (TC-BGM)~\cite{tcbgm} to propagate the
electronic wave function in the field of the classically moving nuclei.
We combine electron removal 
probabilities in an impact parameter by impact parameter fashion for 
three perpendicular orientations of the dimer with respect to the
rectilinear projectile trajectories
and then
orientation-average the results to calculate absolute yields, i.e., cross sections, 
for the processes of interest. 
As it turns out, it is crucial to describe the O$^{3+}$ ion by a screened Coulomb potential
and take into account that its $2s$ subshell is occupied.

Our model is explained in Sec.~\ref{sec:model}. 
In Sec.~\ref{sec:results} we present and discuss our results in comparison with the
experimental data and the previous COBM results.
The paper ends with a few concluding remarks in Sec.~\ref{sec:conclusions}.
Atomic units, characterized by $\hbar=m_e=e=4\pi\epsilon_0=1$, are used unless otherwise stated.

\section{Model}
\label{sec:model}
The basic assumptions of our theoretical model are that (i) the 
projectile ion travels on a 
straight-line classical trajectory with constant speed $v$ (semiclassical approximation), 
and (ii) the target system can be
described as two independent atoms, fixed in space during the collision at a
distance that corresponds to the equilibrium bond length $R_e$ of the neon dimer. We use
the value $R_e=5.86$ a.u.~\cite{Cybulski99, Iskandar15}.
Following the work of, e.g., L\"uhr and Saenz for collisions involving 
H$_2^+$ \cite{Luehr09} and H$_2$ \cite{Luehr10} we consider three perpendicular orientations 
of the target with respect to the projectile path: In orientation I, the dimer is aligned 
parallel to the projectile beam axis. In orientation II it is perpendicular to the
projectile beam in the scattering plane, while in orientation III it is
perpendicular to the scattering plane (see Fig.~1 of Ref.~\cite{Luehr09} for a sketch
of the geometry). 
We calculate electronic transition probabilities for the 
processes of interest as functions of the (scalar) impact parameter $b$, 
measured with respect to the center-of-mass of the dimer, for these
three orientations and construct an
orientation-average for each process $j$ according to  
\begin{equation}
	P_j^{\rm ave} (b) = \frac{1}{3}\left( P_j^{\rm I}(b) +  P_j^{\rm II}(b) + P_j^{\rm III}(b) \right) .
	\label{eq:pave}
\end{equation}
This orientation-averaged probability is then integrated over the impact parameter to calculate the
cross section
\begin{equation}
	\sigma_j^{\rm ave} = \int P_j^{\rm ave}(b) d^2 b = 2\pi \int_0^\infty bP_j^{\rm ave}(b) db .
	\label{eq:tcs}
\end{equation}

In the following subsection we describe how the ion-atom problem is solved. The subsequent
Sec.~\ref{sec:model-analysis} deals with the combination of the ion-atom results to 
obtain probabilities and cross sections for the ion-dimer system.

\subsection{Ion-atom collision calculations}
\label{sec:model-atom}
The ion-atom collision calculations are carried out at the level of the independent electron
model (IEM), i.e., electron-correlation effects are neglected and the Hamiltonian is
assumed to have one-body character, ${\hat H} (t) \approx \sum_{i} {\hat h}_i(t)$,
with a single-electron Hamiltonian of the form
\begin{equation}
	\hat h(t) = -\frac{1}{2} \nabla^2 + v_{T}(r) + v_{P}({\bf r},t) ,
	\label{eq:heff}
\end{equation}
such that the many-electron time-dependent Schr\"odinger equation separates into a set of
single-particle equations for the initially populated orbitals.
In Eq.~(\ref{eq:heff})
$v_{T}$ denotes a spherically-symmetric effective target potential which includes the nuclear Coulomb potential
(with charge number $Z_T=10$ for Ne) and ground-state Hartree screening and exchange
potentials obtained from the optimized potential method (OPM) of density functional theory
(DFT)~\cite{ee93}.
As a consequence of the exact treatment of exchange effects in the OPM, $v_T$ falls off like $-1/r$
asymptotically and the exchange potential obtained from the numerical solution of the OPM
integral equation exhibits a structure at intermediate $r$ 
which can be interpreted as a manifestation of the shell structure of the atom~\cite{ee99b}.

The projectile potential $v_{P}$ is a bare Coulomb potential with charge number $Z_P=3$ for
Li$^{3+}$ projectiles and a screened Coulomb potential of Green-Sellin-Zachor~\cite{Green69} 
form for O$^{3+}$:
\begin{equation}
	v_{P}({\bf r},t) = v_P(r_P) = - \frac{1}{r_P} \left[ \frac{5}{1+H(e^{r_P/d}-1)}+3\right]   .
	\label{eq:vpgsz}
\end{equation}
In Eq.~(\ref{eq:vpgsz}) $r_P = |{\bf r}-{\bf R}(t)|$ is the distance between the active electron and
the projectile nucleus, whose position vector follows the straight-line path ${\bf R}(t)=(\tilde b,0,v t)$
where $\tilde b$ is the impact parameter with respect to the target atom [to be distinguished from
the impact parameter $b$ in Eqs.~(\ref{eq:pave}) and (\ref{eq:tcs})]. 
The parameters $d=0.476$ and $H=3.02d$,  
taken from Table I of Ref.~\cite{Szydlik74}, were 
determined by a modified Hartree-Fock procedure described in that paper.
The potential (\ref{eq:vpgsz}) interpolates between $-3/r_P$ for long and 
$-8/r_P$ for short distances, as it should from the perspective of an (active) electron 
placed on the target atom initially and ionized or captured by the projectile during the
course of the collision.
One can view our choice of Hamiltonian (\ref{eq:heff}) as a no-response approximation
to a full DFT treatment of the problem in which time-dependent
electron-electron interaction effects are neglected~\cite{tom98, hjl18}.

The eight Ne $L$-shell electrons are propagated subject to the Hamiltonian (\ref{eq:heff})
using a basis representation obtained from the TC-BGM, while the $K$-shell electrons
are assumed to be passive. 
The $K$-shell electrons of the  O$^{3+}$ projectile ion are assumed to be passive as well,
whereas the projectile $L$-shell electrons have to be treated with more care, as is
explained further below.

The basis used includes the $2s$ to $4f$ target orbitals and all projectile orbitals
from $1s$ ($2s$) to $7i$ for Li$^{3+}$ (O$^{3+}$), augmented by electron-translation
factors to ensure Galilean invariance.
We use atomic orbitals with real instead of the standard complex
spherical harmonics as their
angular parts. This has the advantage that all basis states have even (`gerade')
or odd (`ungerade') symmetry with respect to reflections about the scattering plane
and do not mix during propagation. We denote these symmetry-adapted orbitals by the quantum numbers
$nlm_g$ and $nlm_u$ in the following.

The target and projectile two-center basis is further augmented by
sets of 35 BGM pseudo states of gerade symmetry and 21 states of ungerade symmetry
constructed in the usual way by operating with powers of a regularized projectile
potential operator on the target eigenstates~\cite{tcbgm}. 
The asymptotic population of these states, when orthogonalized to the target and projectile
two-center basis, can be interpreted as direct transfer to the continuum.
Calculations have been carried
out from an initial to a final projectile--target distance of 50 a.u.
to ensure asymptotic convergence below the one-percent level and
on fine impact-parameter grids to 
resolve the rich structure at the impact energy
of $E=2.81$ keV/amu (corresponding to $v=0.335$ a.u.), which was used in the
experiment~\cite{Iskandar15}. 
More details about the solution of the ion-atom collision problem using the TC-BGM 
are provided in Refs.~\cite{tcbgm, hjl18}.

\begin{figure}
\vspace{-8\baselineskip}
\begin{center}
\resizebox{0.8\textwidth}{!}{\includegraphics{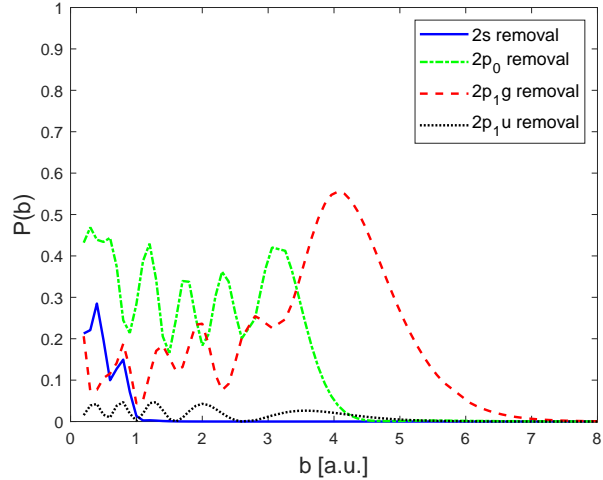}}  
\vspace{-7\baselineskip}
\caption{Single-particle probabilities for electron removal from the Ne $L$ shell by 2.81 keV/amu
Li$^{3+}$ impact plotted as functions of the impact parameter. 
}
\label{fig1}
\end{center}
\end{figure}

Figure~\ref{fig1} shows the single-electron removal probabilities, obtained by
exploiting the unitarity of the coupled-channel problem and subtracting
the asymptotic target orbital populations from unity, for the Li$^{3+}$ projectile. The probabilities
are almost indistinguishable from the single-electron capture probabilities, i.e., direct
transfer to the continuum is negligible (less than 0.5\%).
Clearly, electron removal is stronger for the initial Ne($2p_0$) and Ne($2p_{1g}$) electrons than for the
$2s$ electrons which are more strongly bound and cannot be captured very efficiently into
hydrogenlike Li$^{2+}$. Qualitatively, this can be understood by comparing the
energy eigenvalues of the relevant target ($\varepsilon_{{\rm Ne}(2s)}^{\rm OPM}=-1.718$ a.u. versus
$\varepsilon_{{\rm Ne}(2p)}^{\rm OPM}=-0.851$ a.u.) and projectile 
($\varepsilon_{{\rm Li}^{2+}(n=2)}=-1.125$ a.u.) orbitals and keeping
in mind that capture to lower-lying states is more likely because of the 
Stark shifts of the target states in the projectile potential. This simple
argument suggests that capture of Ne($2p$) electrons to projectile states
of principal quantum number $n=2$ is the
strongest channel and indeed this is what the numerical calculations show.
The removal of the Ne($2p_{1u}$) electrons is relatively weak, since fewer final states
are available in the ungerade symmetry case.
A more detailed analysis would require to compute correlation diagrams and
quasimolecular couplings.

For O$^{3+}$ impact the situation is complicated by the fact that Pauli blocking may prevent 
some electron capture transitions. As mentioned above, we consider both the Ne and the O$^{3+}$ $K$-shell
electrons as passive and do not include those states in the TC-BGM basis. This is justified by the
large binding energies of the $K$-shell electrons and the weak couplings of the $1s$ states
to other basis states.
Such an approach
does not work for the occupied $L$ shells as some state-to-state couplings are
strong and simply eliminating occupied states from the coupled-channel calculations 
contaminates some of the open channels. 
To illustrate these points, we note that
in a TC-BGM calculation with the full basis the single-particle transfer probability from Ne($2s$) to
O$^{3+}(2s)$ becomes very close to unity at some impact parameters, while test calculations in
which the (occupied) O$^{3+}(2s)$ state was removed from the basis
resulted in sizable transfer to the continuum---a process
that should be ineffective at low collision energy.

In order to deal with this situation we subtracted the single-particle probabilities for
the transitions Ne$(2l) \rightarrow $ O$^{3+}(2s)$ from the Ne$(2l)$ electron removal 
probabilities and interpreted the results as the `true' removal probabilities. 
This seemingly naive procedure can be justified based on the principle of detailed balance
[which asserts that the probability for a transition from, say, Ne$(2s)$ to O$^{3+}(2s)$ 
equals the probability for a transition from O$^{3+}(2s)$ to Ne$(2s)$] and the
inclusive probability formalism of Ref.~\cite{hjl85}. The argument is presented in the
Appendix.

We note that we ignored Pauli blocking due to the presence of one $2p$ electron
in O$^{3+}$ based on the rationale that this should be a weak effect given that five out of six 
states in the $2p$ subshell are vacant.

\begin{figure}
\vspace{-8\baselineskip}
\begin{center}
\resizebox{0.8\textwidth}{!}{\includegraphics{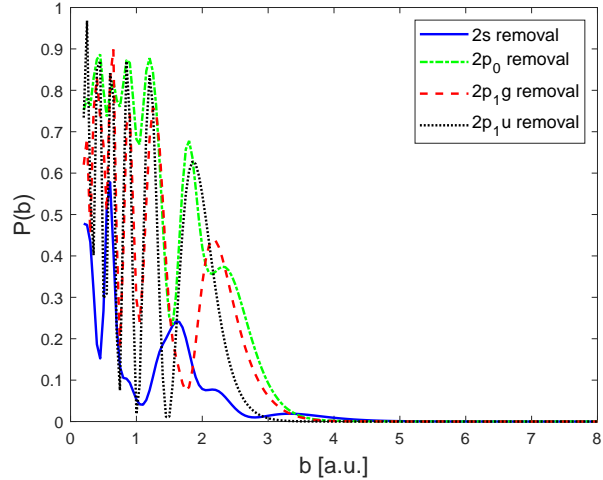}}  
\vspace{-7\baselineskip}
\caption{Single-particle probabilities for electron removal from the Ne $L$ shell by 2.81 keV/amu
O$^{3+}$ impact plotted as functions of the impact parameter. The probabilities are
corrected for the presence of the projectile $2s$ electrons as described in the text.
}
\label{fig2}
\end{center}
\end{figure}

The resulting single-particle electron removal probabilities for O$^{3+}$-Ne collisions are
presented in Fig.~\ref{fig2}. Similarly to those of the Li$^{3+}$-Ne system (cf.~Fig.~\ref{fig1})
they show rich structure as a function of impact parameter, but the details are quite different.
Notably, all probabilities reach higher values, not far from unity for the Ne$(2p)$ initial states and
up to 0.6 for Ne$(2s)$, the latter to be contrasted with a maximum 
removal probability $p_{2s}^{\rm rem}$
of approximately 0.3 for Li$^{3+}$ impact. 
Also, $p_{2s}^{\rm rem}$ extends to significantly larger impact parameters for O$^{3+}$ 
than for Li$^{3+}$ projectiles, while the trend is opposite for $2p$ removal. 
The main reason for the increased probabilities in the $0< b \lessapprox 2.5$ a.u. range 
is the lower energy eigenvalue of the
(vacant)  O$^{3+}(2p)$ orbitals
at -1.868 a.u. as compared to -1.125 a.u. for hydrogenlike Li$^{2+}(2p)$, which makes capture
(from all states) more effective.
The increased Ne$(2s)$-vacancy production probability will become important for the role of ICD to be
discussed in the next section. 

\subsection{Analysis of electronic processes resulting in ICD, CE, and RCT}
\label{sec:model-analysis}
We now look at the neon dimer in each of the three orientations described above and combine
ion-atom probabilities in an impact parameter by impact parameter fashion to calculate the probabilities
on the right hand side of Eq.~(\ref{eq:pave}) for the three processes of interest. 
For orientation I in which the dimer is parallel to
the ion beam axis the situation is simple, since the impact parameters with respect to both atoms are
the same and coincide with the impact parameter with respect to the center of mass of the dimer,
i.e., $\tilde b \equiv b_{\rm I} = b$.

For each value of $b$ considered, we proceed by determining the corresponding atomic impact parameters
for orientations II and III and then carry out TC-BGM calculations at those impact parameters to
avoid interpolations when combining and orientation-averaging probabilities  
for the ion-dimer system.
For orientation III in which the dimer is perpendicular to the scattering plane both atomic 
impact parameters are the same and are given by $b_{\rm III} = \sqrt{b^2 + (R_e/2)^2}$. For orientation II
the two atomic impact parameters are different. The one with respect to the closer atom
is $b_{\rm II}^{(1)}=|(R_e/2)-b|$ and the other one is $b_{\rm II}^{(2)}=b+(R_e/2)$.

As mentioned in the Introduction 
ICD, CE, and RCT can be associated with specific one- and two-electron removal 
processes~\cite{Iskandar15, Cassimi19}.
We calculate these processes by combining all products of single-particle
probabilities which contribute to a given outcome, i.e., by
a straightforward multinomial analysis 
of the combined ion--two-atom system. 
This approach corresponds to an IEM in which in addition to electron correlations the effects of the
Pauli exclusion principle are neglected as well~\cite{McGuire97, santanna98, tom98}. 
It has been used in a large number of theoretical
studies of ion-atom and ion-molecule collision problems (see, e.g., 
Refs.~\cite{montanari12, tom14, leung17, terekhin18} for recent examples). 

Let us exemplify the procedure for 
the simplest case of orientation I in which both atomic impact parameters are the same.
The probability for finding one vacancy in one of the Ne($2s$) orbitals is given by
\begin{equation}
	P_{2s^{-1}}^{\rm I}(b) =
	4 p_{2s}^{\rm rem}(b_{\rm I}) (1-p_{2s}^{\rm rem}(b_{\rm I}))^3 (1-p_{2p_0}^{\rm rem}(b_{\rm I}))^4 
	(1-p_{2p_{1g}}^{\rm rem}(b_{\rm I}))^4 (1-p_{2p_{1u}}^{\rm rem}(b_{\rm I}))^4 ,
	\label{eq:picd}
\end{equation}
where $b=b_{\rm I}$. The $(1-p_{2s,2p}^{\rm rem})$ terms in this expression account for the requirement 
that all $2p$ electrons and three out of four $2s$
electrons of the two atoms remain bound. The multiplication factor of four arises because each of the
four initial $2s$ electrons can be the one that is removed.
The $2s$-vacancy process (\ref{eq:picd}) can be associated with ICD.

Similarly, the probability for the removal of one $2p$ electron from each atom is given by
\begin{eqnarray}
	P_{2p^{-1}, 2p^{-1}}^{\rm I} &=&
	(1-p_{2s}^{\rm rem})^4 [2p_{2p_0}^{\rm rem}(1-p_{2p_0}^{\rm rem})
	(1-p_{2p_{1g}}^{\rm rem})^2 (1-p_{2p_{1u}}^{\rm rem})^2 
	+ 2p_{2p_{1g}}^{\rm rem}(1-p_{2p_{1g}}^{\rm rem})
	\nonumber \\
	&& \mbox{} \times
	(1-p_{2p_{0}}^{\rm rem})^2 (1-p_{2p_{1u}}^{\rm rem})^2
	 +  2p_{2p_{1u}}^{\rm rem}(1-p_{2p_{1u}}^{\rm rem})
	(1-p_{2p_{0}}^{\rm rem})^2 (1-p_{2p_{1g}}^{\rm rem})^2 ]^2 ,
	\label{eq:pce}
\end{eqnarray}
where we have omitted the impact parameter dependence for ease of notation. 
The first factor involving $p_{2s}^{\rm rem}$ ensures that no inner-valence vacancy is
created. The three terms in square brackets account for the removal of one 
electron from either the $2p_0$, the $2p_{1g}$, or the $2p_{1u}$ orbital and the whole
expression is squared to ensure that one-electron removal happens on both atoms
simultaneously (and independently). 
The probability (\ref{eq:pce}) can be associated with CE.

It was argued in Refs.~\cite{Iskandar15, Iskandar18} that double $2p$ removal from one atom may result
in the third observed process, RCT, but not necessarily so, since the system can also 
dissociate as is, giving rise to one doubly-charged and one neutral fragment. 
The experiment was blind to the latter channel and in the COBM calculations
reported along with the measurements it was assumed that 50\% 
of double removal from one atom will lead to RCT while the other 50\% result in
Ne$^{2+}$ + Ne production~\cite{Iskandar15}.

Within the IEM, 
removing two $2p$ electrons from one atom while the other atom remains intact is represented by
\begin{eqnarray}
	P_{2p^{-2}}^{\rm I} &=&
	2(1-p_{2s}^{\rm rem})^4 
	[(p_{2p_0}^{\rm rem})^2 
	(1-p_{2p_{1g}}^{\rm rem})^2 (1-p_{2p_{1u}}^{\rm rem})^2
	+ (p_{2p_{1g}}^{\rm rem})^2 
	(1-p_{2p_{0}}^{\rm rem})^2 (1-p_{2p_{1u}}^{\rm rem})^2 
	\nonumber \\
	&& \mbox{} 
	 +  (p_{2p_{1u}}^{\rm rem})^2 
	(1-p_{2p_{0}}^{\rm rem})^2 (1-p_{2p_{1g}}^{\rm rem})^2 
	+ 2 p_{2p_0}^{\rm rem} (1-p_{2p_0}^{\rm rem}) 2p_{2p_{1g}}^{\rm rem} (1-p_{2p_{1g}}^{\rm rem}) 
          (1-p_{2p_{1u}}^{\rm rem})^2	
	  \nonumber \\
	  && \mbox{}
	  + 2 p_{2p_{0}}^{\rm rem} (1-p_{2p_{0}}^{\rm rem}) 2p_{2p_{1u}}^{\rm rem} (1-p_{2p_{1u}}^{\rm rem}) 
          (1-p_{2p_{1g}}^{\rm rem})^2
          + 2 p_{2p_{1g}}^{\rm rem} (1-p_{2p_{1g}}^{\rm rem}) 2p_{2p_{1u}}^{\rm rem} (1-p_{2p_{1u}}^{\rm rem}) 
	  (1-p_{2p_{0}}^{\rm rem})^2]
	  \nonumber \\
	  && \mbox{}
	  \times [(1-p_{2p_{0}}^{\rm rem})^2 (1-p_{2p_{1g}}^{\rm rem})^2 (1-p_{2p_{1u}}^{\rm rem})^2  ] .
	  \label{eq:prct}
\end{eqnarray}
While this expression is lengthy, the interpretation of each term is straightforward. 
The first square bracket 
accounts for the removal of
two electrons from one of the atoms from either the same $2p$ orbital or from two different orbitals, 
the latter terms being multiplied
by two factors of two to account for the fact that both electrons in a given orbital are equally likely to be
removed or not. The expression in the second square bracket takes care of the requirement 
that no $2p$ electron be removed from the second atom and the overall prefactor of two
is there since it can be one or the other atom that is ionized.
If one rearranges the terms in Eq.~(\ref{eq:prct}) and
compares the whole expression with Eq.~(\ref{eq:pce}) one obtains
\begin{eqnarray}
	P_{2p^{-1}, 2p^{-1}}^{\rm I} - P_{2p^{-2}}^{\rm I} &=&
	2(1-p_{2s}^{\rm rem})^4 (1-p_{2p_{0}}^{\rm rem})^2 (1-p_{2p_{1g}}^{\rm rem})^2 
		 (1-p_{2p_{1u}}^{\rm rem})^2 [
		(p_{2p_0}^{\rm rem})^2 (1-p_{2p_{1g}}^{\rm rem})^2 (1-p_{2p_{1u}}^{\rm rem})^2
		\nonumber \\
	&& \mbox{}	+ (p_{2p_{1g}}^{\rm rem})^2 (1-p_{2p_{0}}^{\rm rem})^2 (1-p_{2p_{1u}}^{\rm rem})^2
		+ (p_{2p_{1u}}^{\rm rem})^2 (1-p_{2p_{0}}^{\rm rem})^2 (1-p_{2p_{1u}}^{\rm rem})^2] \ge 0 ,
\end{eqnarray}
i.e., the prediction that CE is stronger than RCT, even if one makes the extreme assumption that 
double removal from one atom will always result in RCT.

\begin{figure}
\vspace{-5\baselineskip}
\begin{center}$
\begin{array}{cc}
\resizebox{0.52\textwidth}{!}{\includegraphics{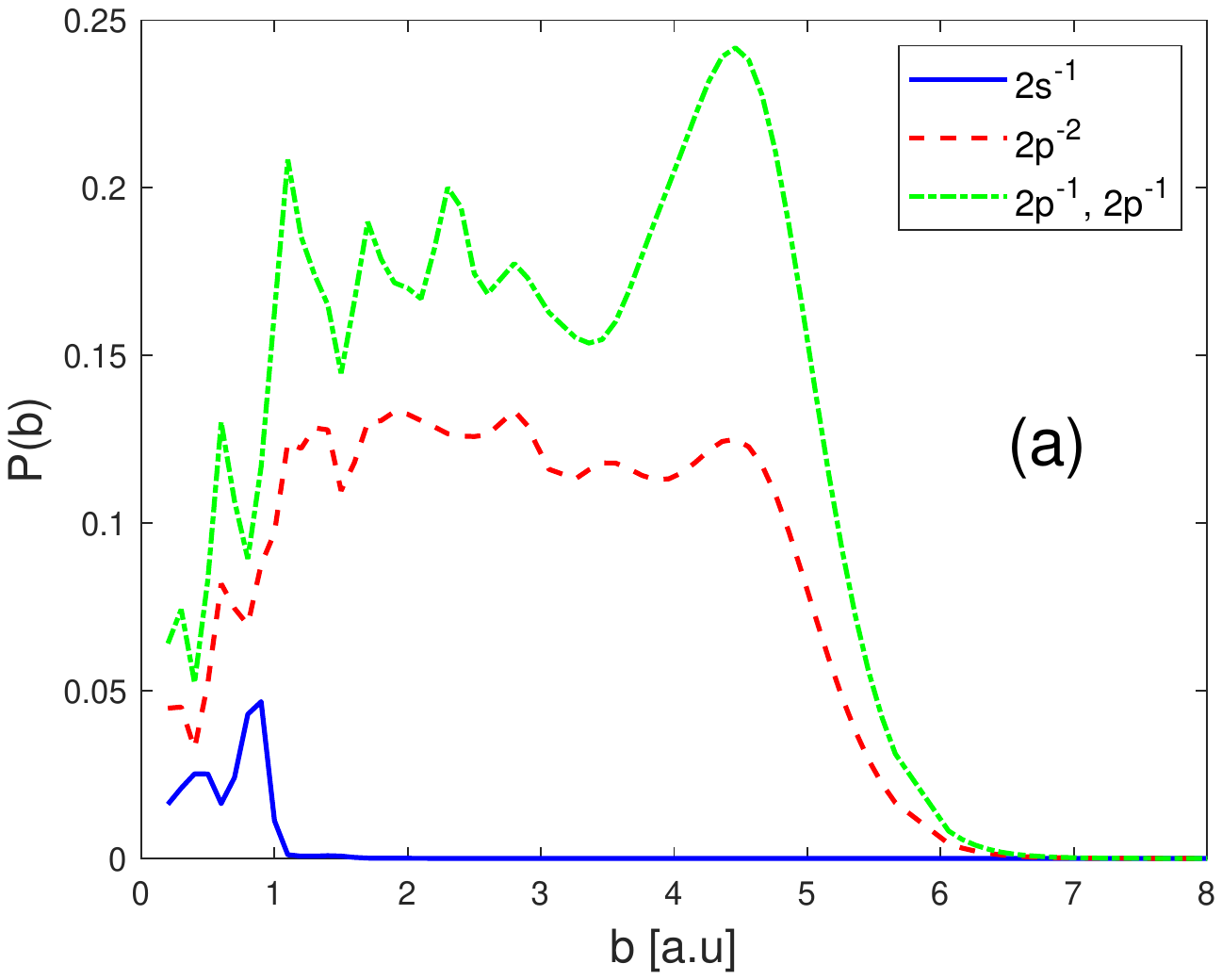}}&
\resizebox{0.52\textwidth}{!}{\includegraphics{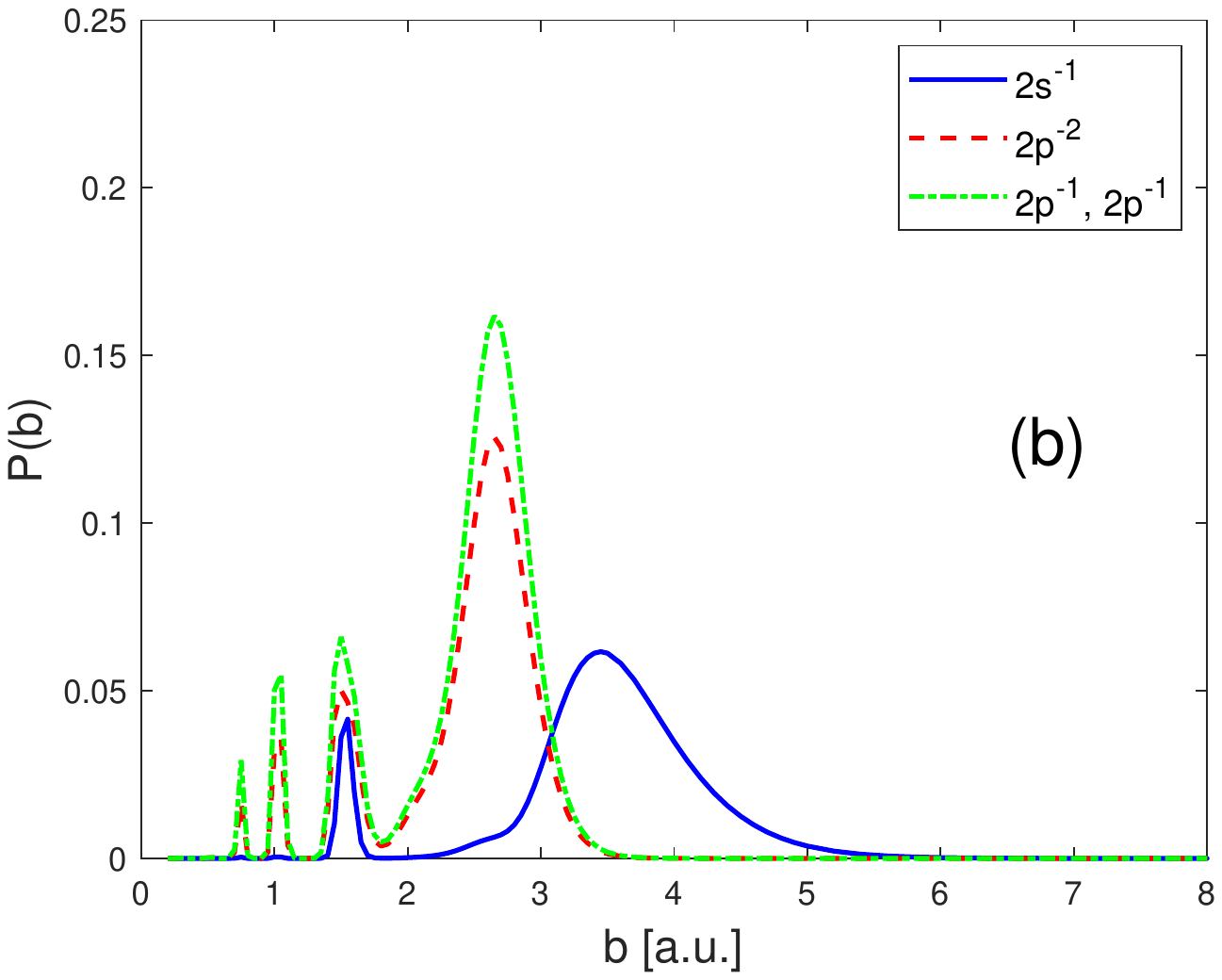}}
\end{array}$
\vspace{-5\baselineskip}
\caption{%
	Probabilities for $2s^{-1}$, $2p^{-2}$, and ($2p^{-1},2p^{-1}$) production 
	in (a) Li$^{3+}$ and
	(b) O$^{3+}$ collisions with Ne$_2$ in orientation I at $E=2.81$ keV/amu.}
\label{fig3}
\end{center}
\end{figure}

This can be seen in Fig.~\ref{fig3} in which 
the probabilities (\ref{eq:picd}) to (\ref{eq:prct}) are plotted as functions of the impact
parameter $b$ for both Li$^{3+}$ [panel~(a)] and 
O$^{3+}$ [panel~(b)] collisions, using the same scales on the $x$ and $y$ axes to
ease the comparison. 

As can be expected from the ion-atom single-particle probabilities shown in Figs.~\ref{fig1}
and~\ref{fig2} the results for the two projectiles are quite different. The $2p$ removal
processes (\ref{eq:pce}) and (\ref{eq:prct}) are significantly stronger for
Li$^{3+}$ than for O$^{3+}$ projectiles and extend to larger impact parameters. The first
part of this observation may seem surprising given that the probabilities displayed
in Fig.~\ref{fig1} (for Li$^{3+}$) tend to be smaller than those of Fig.~\ref{fig2} (for O$^{3+}$). 
However, one has
to keep in mind that both Eqs.~(\ref{eq:pce}) and (\ref{eq:prct}) 
include factors of the type $(1-p_{2p}^{\rm rem})$ which correspond to the fact 
that ten out of twelve $2p$ electrons are not removed. 
These factors act as effective
suppression factors when the single-particle probabilities approach unity.

For the $2s$-vacancy production (\ref{eq:picd}) the situation is reversed and the
O$^{3+}$ projectile is overall more effective than Li$^{3+}$. 
Again, it is a consequence of the $(1-p_{2p}^{\rm rem})$ factors that
the shallow maximum of the O$^{3+}$-impact $2s$ single-particle probability around $\tilde b \approx 3.3$ a.u.
(cf. Fig.~\ref{fig2}) results in the main peak of $P_{2s^{-1}}^{\rm I}$, while the process is mostly
suppressed at smaller impact parameters.

\begin{figure}
\vspace{-5\baselineskip}
\begin{center}$
\begin{array}{cc}
\resizebox{0.52\textwidth}{!}{\includegraphics{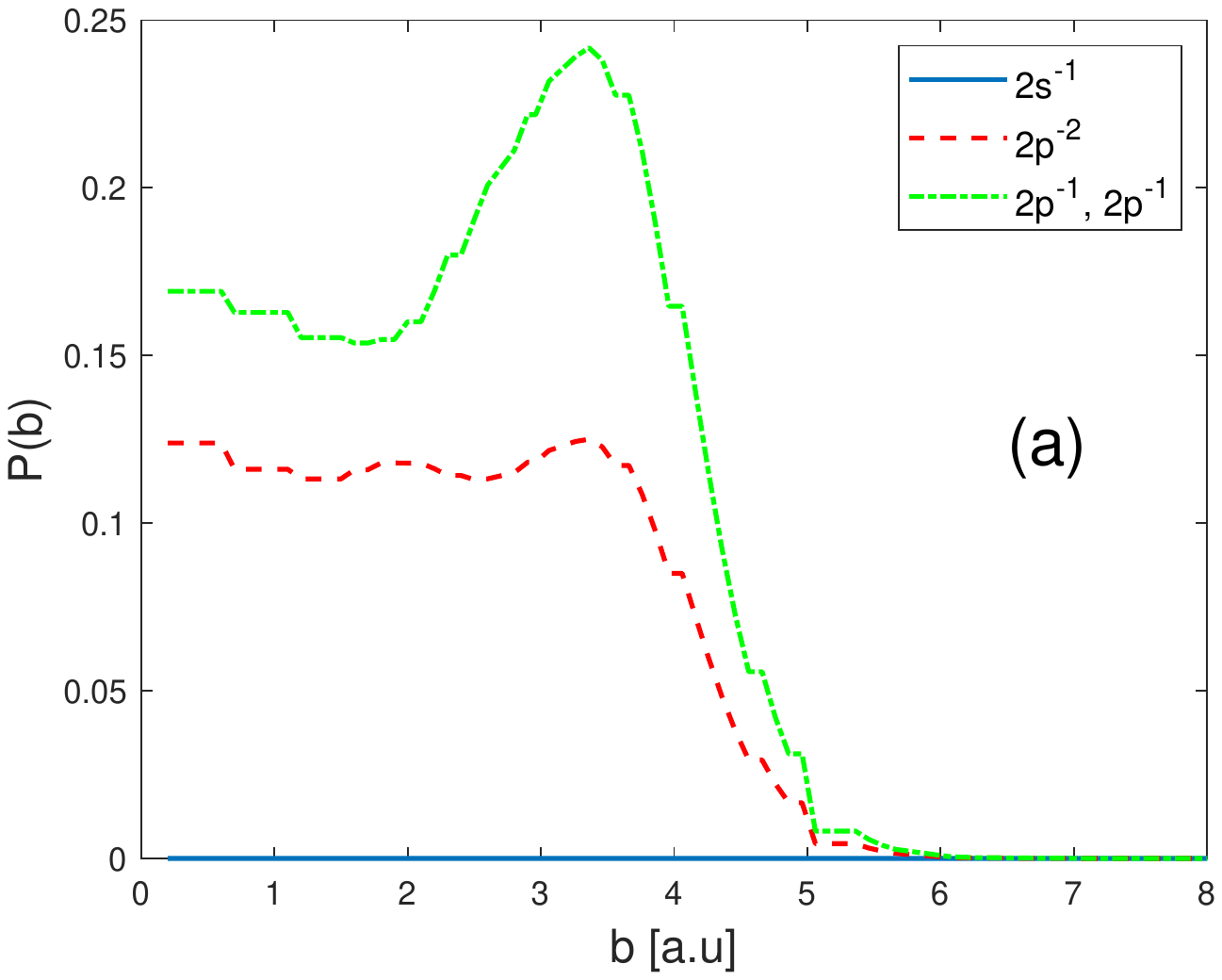}}&
\resizebox{0.52\textwidth}{!}{\includegraphics{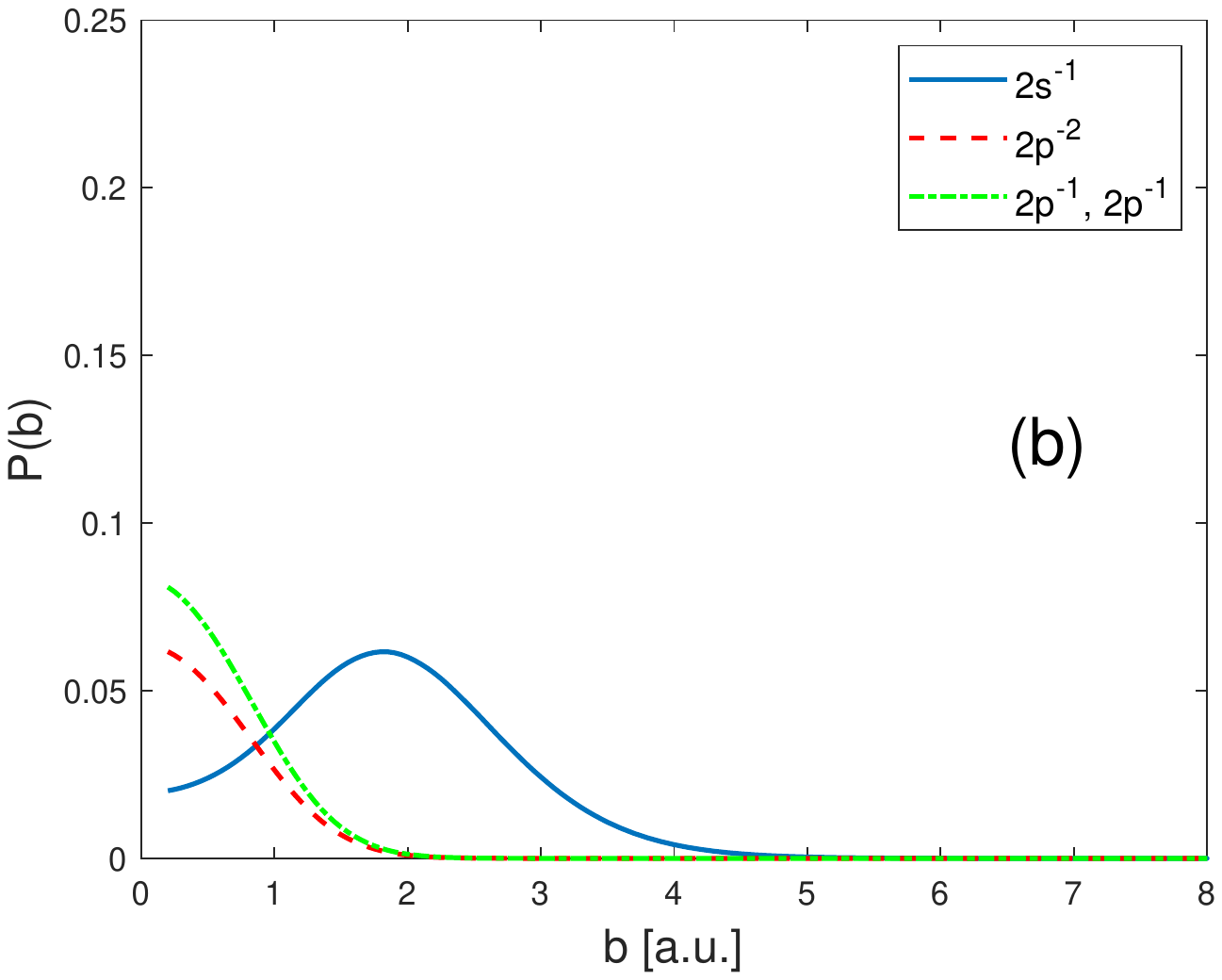}}
\end{array}$
\vspace{-5\baselineskip}
\caption{%
	Probabilities for $2s^{-1}$, $2p^{-2}$, and ($2p^{-1},2p^{-1}$) production 
	in (a) Li$^{3+}$ and
	(b) O$^{3+}$ collisions with Ne$_2$ in orientation III at $E=2.81$ keV/amu.}
\label{fig4}
\end{center}
\end{figure}

For orientation III one can summarize the situation as follows: 
The expressions (\ref{eq:picd})--(\ref{eq:prct}) remain
unchanged except that the impact parameters on the left and right hand sides are now different, i.e.,
$b\neq b_{\rm III} = \sqrt{b^2 + (R_e/2)^2}$. 
One then sees (in Fig.~\ref{fig4}) the probability distributions which occur at impact parameters 
$b\ge R_e/2$ in
orientation I at smaller impact parameters and stretched out over a wider interval.
The structures occuring at $b< R_e/2$ in orientation I are eliminated from the plot
for orientation III, since the projectile
never gets close enough to the two atoms. This is why the $2s$-vacancy
production process is absent for Li$^{3+}$ projectiles [Fig.~\ref{fig4}(a)].

\begin{figure}
\vspace{-5\baselineskip}
\begin{center}$
\begin{array}{cc}
\resizebox{0.52\textwidth}{!}{\includegraphics{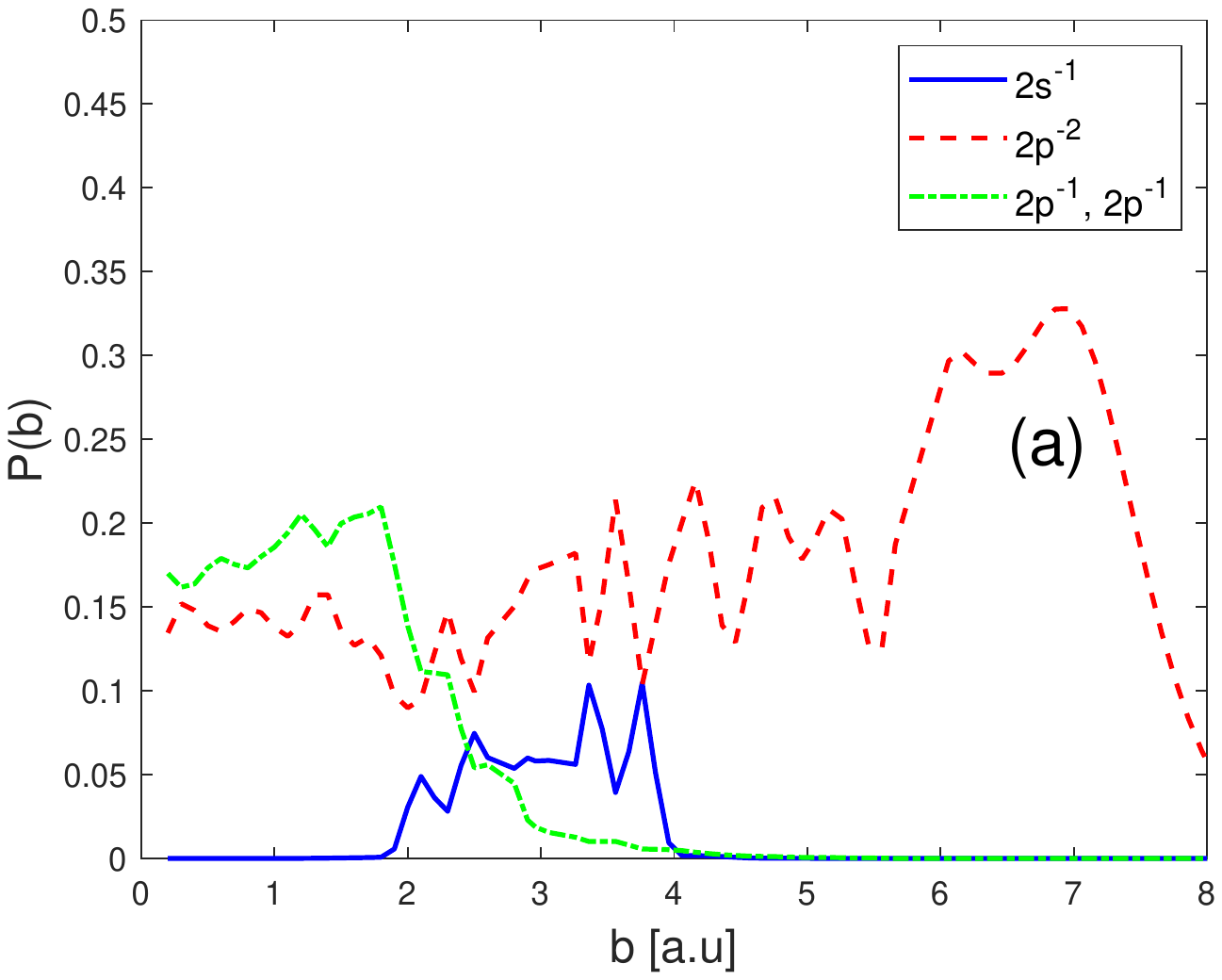}}&
\resizebox{0.52\textwidth}{!}{\includegraphics{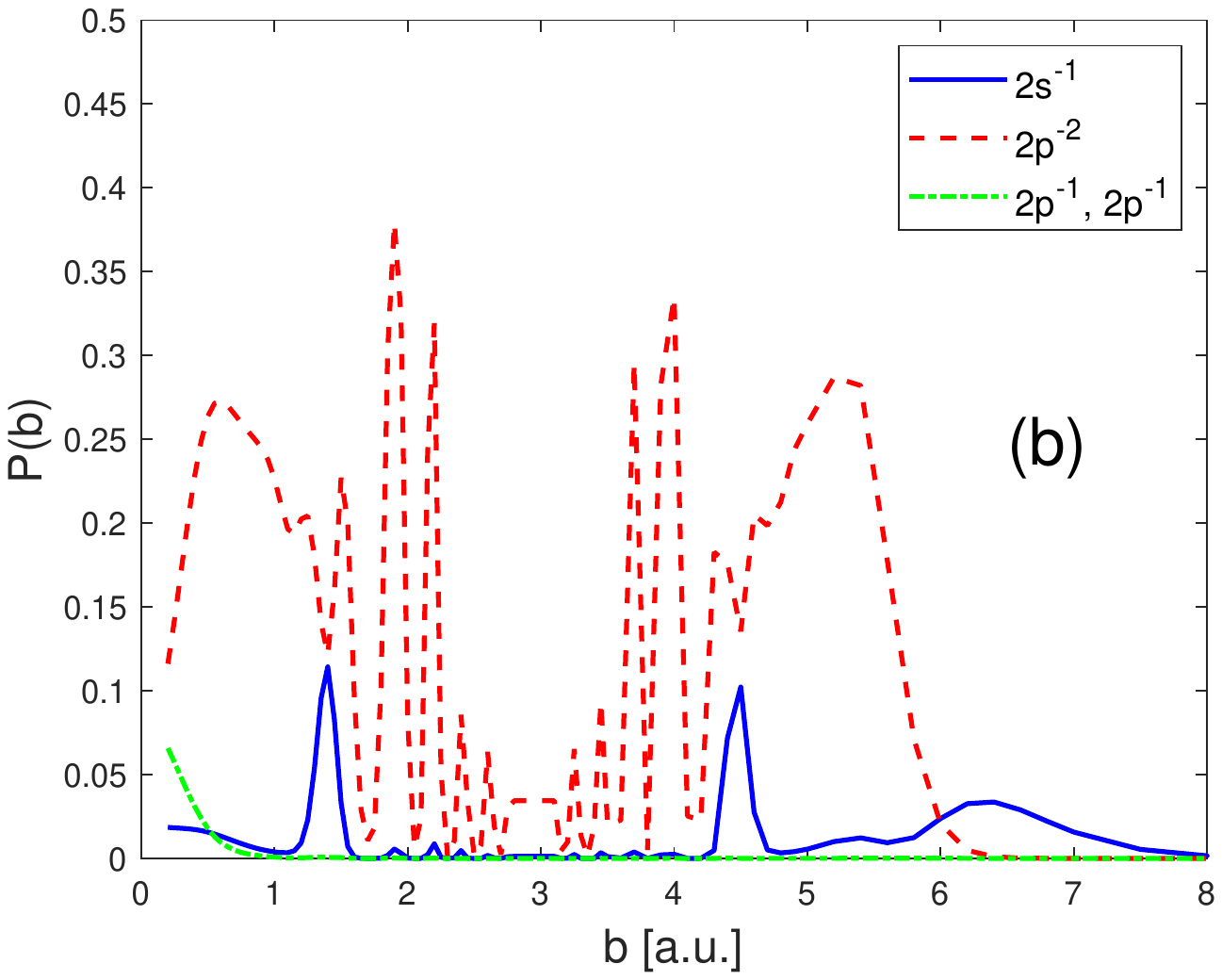}}
\end{array}$
\vspace{-5\baselineskip}
\caption{%
	Probabilities for $2s^{-1}$, $2p^{-2}$, and ($2p^{-1},2p^{-1}$) production 
	in (a) Li$^{3+}$ and
	(b) O$^{3+}$ collisions with Ne$_2$ in orientation II at $E=2.81$ keV/amu.}
\label{fig5}
\end{center}
\end{figure}

Orientation II 
in which the two atomic impact parameters are not the same, 
produces lengthier (but still straightforward) mathematical expressions and more complicated patterns 
for the three processes. 
This orientation does allow for
$P_{2p^{-2}} > P_{2p^{-1}, 2p^{-1}}$ and in a quite
pronounced way, in particular for O$^{3+}$ projectiles as shown in Fig.~\ref{fig5}(b).
For this projectile the $P_{2p^{-1}, 2p^{-1}}$ probability is essentially zero except at $b<1$ a.u.,
which can be understood by once again
inspecting Fig.~\ref{fig2} and noticing that all $2p$ electron removal probabilities are small
at atomic impact parameters larger than $R_e/2=2.93$ a.u. and are approaching zero rapidly toward more
distant collisions.
Given that both atoms need to be ionized for this process to occur and
the farther atom in this orientation is at least a distance of $R_e/2$ away from
the projectile, $P_{2p^{-1}, 2p^{-1}}$ is very small.
By contrast, $P_{2p^{-2}}$ reaches substantial values, since both electrons can be efficiently
removed from the closer atom. In this case, the obtained distribution is basically symmetric
with respect to $b=R_e/2$ which corresponds to a head-on ion-atom collision.
The same is true for the $2s$-vacancy production process, except that
the shallow peak around $b\approx 6.4$ a.u. is too far out to have a mirror image at small
impact parameters.

For Li$^{3+}$ projectiles [Fig.~\ref{fig5}(a)] the situation is different since the $2p$ single-particle 
removal probabilities extend beyond $b\approx R_e/2$ (cf. Fig.~\ref{fig1}) and more substantial overlaps
between contributions from the close and the far atom occur. The $2s$-vacancy production channel
contributes in the interval $2\lessapprox b \lessapprox 4$ a.u., as can be expected from
Fig.~\ref{fig1} and Fig.~\ref{fig3}(a): Only the closer atom can provide a nonzero $p_{2s}^{\rm rem}$ factor 
and it does so only when the atomic impact parameter is one atomic unit or smaller.
The distribution is not symmetric about $b=R_e/2$ because of the contributions
from the $(1-p_{2p}^{\rm rem})$ factors from both atoms.

\begin{figure}
\vspace{-5\baselineskip}
\begin{center}$
\begin{array}{cc}
\resizebox{0.52\textwidth}{!}{\includegraphics{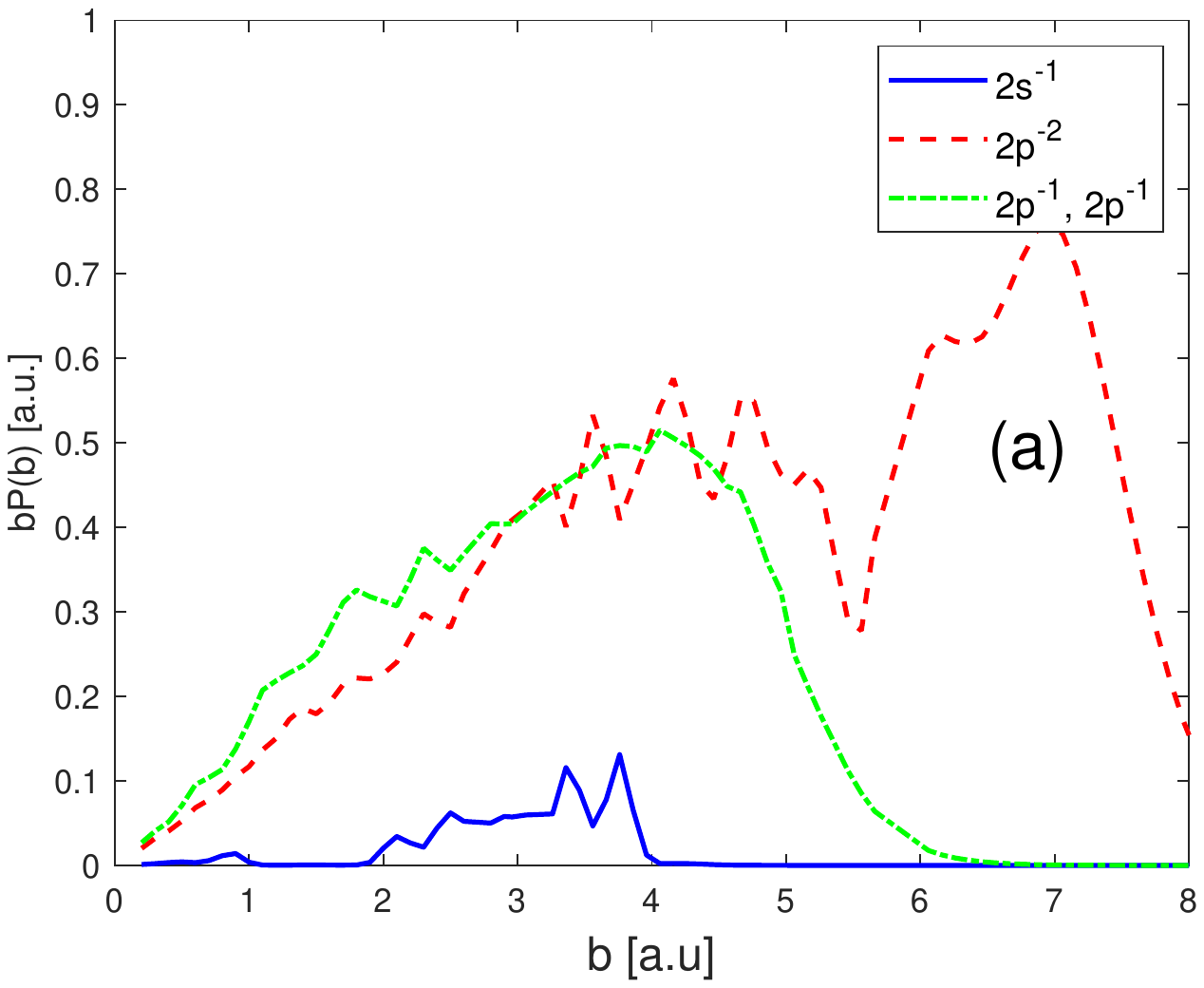}}&
\resizebox{0.52\textwidth}{!}{\includegraphics{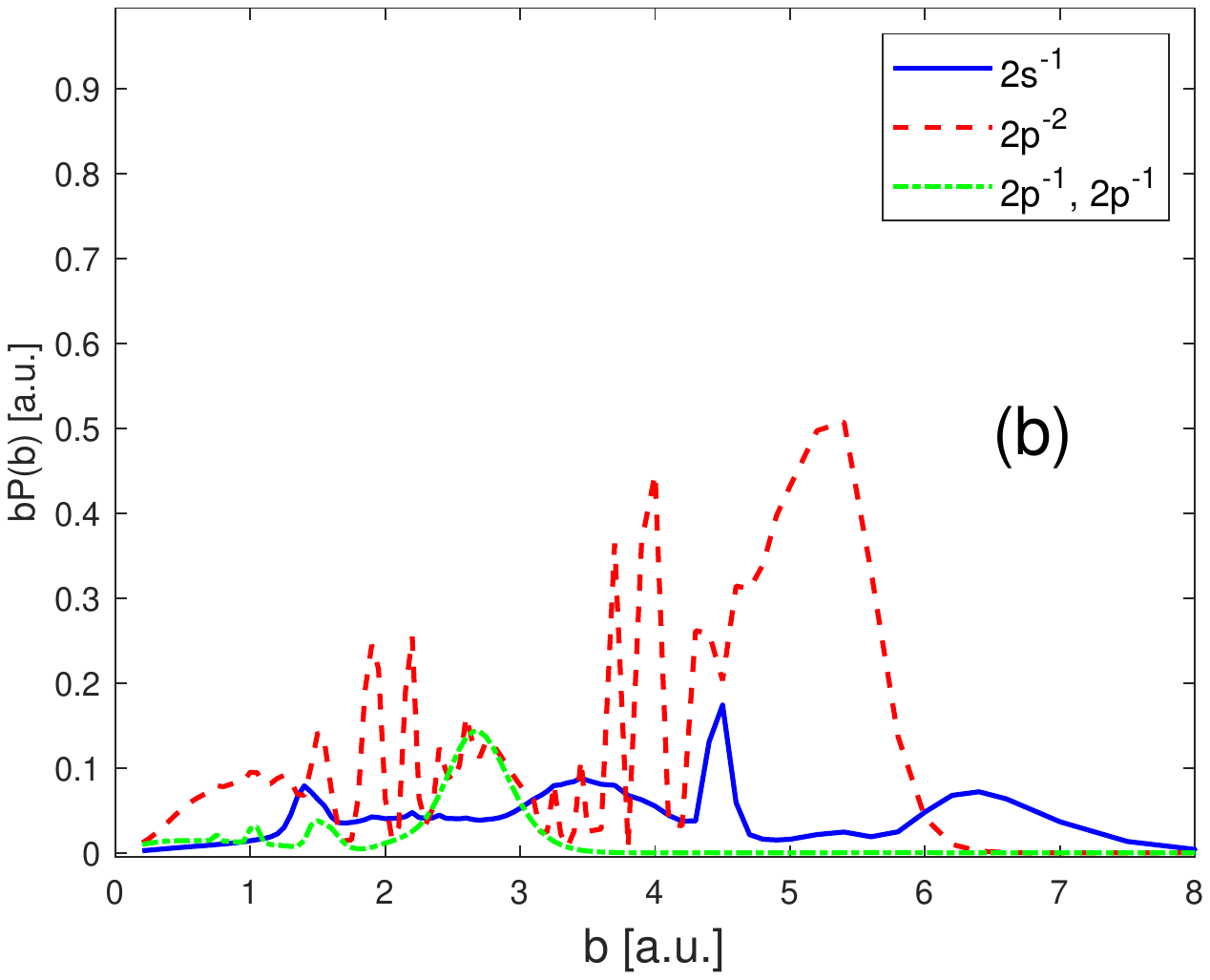}}
\end{array}$
\vspace{-5\baselineskip}
\caption{%
        Orientation-averaged impact-parameter-weighted probabilities 
        for $2s^{-1}$, $2p^{-2}$, and ($2p^{-1},2p^{-1}$) production 
	in (a) Li$^{3+}$ and
	(b) O$^{3+}$ collisions with Ne$_2$ at $E=2.81$ keV/amu.}
\label{fig6}
\end{center}
\end{figure}

Figure~\ref{fig6} displays the orientation-averaged probabilities 
weighted by the impact parameter, i.e., the integrands of the cross section formula (\ref{eq:tcs})
for the three processes of interest.
The most obvious differences between the plots for both projectiles 
are that (i) both $2p$ removal processes are stronger for Li$^{3+}$ [panel~(a)] than for O$^{3+}$ 
[panel~(b)] and, (ii) on a relative scale,
the $2s$-vacancy production process and the
process in which one $2p$ electron is removed from each atom 
switch roles: The former is by far the weakest channel for Li$^{3+}$, while the latter shows
less area under the curve for O$^{3+}$, i.e., a smaller total cross section.

\section{Comparison with experimental and COBM data and discussion}
\label{sec:results}
We now discuss the relative yields obtained 
from comparing the total cross sections
for the three processes.
In order to compare the present results with the 
experimental data for ICD, CE, and RCT 
and the COBM calculations of Ref.~\cite{Iskandar15} we apply the same correction as 
used in that work, namely we assume that only 50\% of $P_{2p^{-2}}$ contributes
to RCT, while 100\% of $P_{2p^{-1},2p^{-1}}$ feeds into CE and 
100\% of $P_{2s^{-1}}$ into the ICD channel. 
The resulting relative yields (in percent) are shown in Table~\ref{tab1}.
For our O$^{3+}$ calculations we also show results obtained from the assumption
that 100\% of $P_{2p^{-2}}$ results in RCT.

\begin{table}
\caption{
\label{tab1}
	Relative yields (in percent) for the three processes of interest. The TC-BGM results marked with a
	star are obtained from the assumption that 100\% of $P_{2p^{-2}}$ contributes to RCT, while in the
	other columns it is assumed that only 50\% contributes to this channel. The COBM and experimental
	data are from Ref.~\cite{Iskandar15}. 
}
\begin{tabular}{lccccc}
\hline \hline
& COBM (Li$^{3+}$) & TC-BGM (Li$^{3+}$) & TC-BGM (O$^{3+}$) & TC-BGM (O$^{3+}$)$^*$ & Expt. \\
\hline
	$2s^{-1}$ (ICD)& 8.0  & 3.6 & 34.7 & 22.7 & 20  \\
$2p^{-1},2p^{-1}$ (CE) & 39.3 & 50.2 & 12.5 & 8.2 & 10  \\
	$2p^{-2}$ (RCT) & 52.7 & 46.2 & 52.8 & 69.1 & 70 \\
\hline \hline
\end{tabular}
\end{table}

Clearly, the calculations for O$^{3+}$ projectiles are in better agreement with the measurements
than those for Li$^{3+}$ impact. In particular, they give the experimentally observed ordering
CE $<$ ICD $<$ RCT, while both the Li$^{3+}$ TC-BGM and the COBM calculation of Ref.~\cite{Iskandar15}
appear to overemphasize the CE channel and underestimate ICD.
These two calculations make different predictions about which one is the strongest channel,
but are nevertheless in fair agreement with each other. 

The fact that the present results
for O$^{3+}$ are in almost perfect agreement with the experimental yields when the `100\% assumption'
is applied to the $2p^{-2}$ channel should perhaps not be overinterpreted given that our model
has several limitations: First, reinspecting Figs.~\ref{fig3} to~\ref{fig5} one observes that the
orientation dependence is quite strong. 
This raises the question whether an orientation-average involving more
than three orientations might yield a different result. While ultimately this can only
be answered by additional calculations we note that in their work for collisions
involving H$_2^+$ \cite{Luehr09} and H$_2$ \cite{Luehr10} L\"uhr and Saenz also found considerable orientation
dependence, but concluded that averages based on the three perpendicular
orientations only were rather accurate.  

Second, the IAM for the dimer coupled with the IEM for the electrons 
of both atoms provides of course
only an approximate framework for the discussion of the collision problem at hand. 
In recent work for a large variety
of multicenter systems, ranging from small covalent molecules to large clusters and biomolecules,
an amended IAM was explored, in which the geometric overlap of effective atomic cross sectional areas was
taken into account~\cite{hjl16, hjl18, hjl20}. 
However, that model has so far only been applied to net cross sections and not to the more detailed
one- and two-electron removal processes studied here.
While such an extension is outstanding, one
can perhaps argue that overall geometric overlap should be small
for a system such as Ne$_2$ whose internuclear distance is large, but that it would affect the three
orientations considered differently 
and would amount to their re-weighting in the 
calculation of the orientation-average. 
To estimate the potential effect 
we applied the extreme assumption that the orientation I
probabilities are to be divided by a factor of two to account for the complete overlap of the atomic cross
sectional areas when viewed from the position of the impinging projectile, 
while the results for the other two orientations remain
unchanged. We found that the relative yields do change, but not very dramatically. In particular, the
ICD yields decrease from 34.7\% and 22.7\% for the two models shown in Table~\ref{tab1} to 34.0\% and 21.6\%.

It is more difficult to estimate the error associated with using the IEM, or, in other words, the effects
of electron correlations, which we have neglected in order to simplify the treatment. While it is known 
that they do play a role in collisional multielectron dynamics~\cite{McGuire97, Belkic08, Aumayr19}, 
it is not clear how they affect the relative yields of interest here.
First-principles many-electron calculations would be required to shed light on this issue.
In their absence, 
we can only say that the fair agreement between our O$^{3+}$ results and the experimental data
does not suggest that electron correlations
are of major importance.

\section{Concluding remarks}
\label{sec:conclusions}
Motivated by a recent joint experimental/theoretical work~\cite{Iskandar15}, 
we have studied specific one- and two-electron removal processes in Li$^{3+}$ and 
O$^{3+}$ collisions with neon dimers at $E=2.81$ keV/amu, representing the target system as
two independent atoms and accounting for the ion-atom electron dynamics 
at the level of the independent electron model. The 
coupled-channel two-center basis generator method has been used to solve the single-particle
Schr\"odinger equation for all active target electrons
taking into account in the O$^{3+}$ case
that the projectile potential is  
of screened Coulomb character
and that the $2s$ subshell is occupied.

We find that the results for both projectiles are markedly different and 
only the O$^{3+}$ calculations yield fair agreement with experimental data for 
ICD, CE, and RCT. In particular, our calculations suggest that ICD
is so weak a process for bare projectiles that it might be hard to measure it.
This is a new piece of information given that the 
classical calculation of Ref.~\cite{Iskandar15} predicted a somewhat higher ICD
yield for Li$^{3+}$ impact.

Together with the conclusion of that paper that ICD can only
be observed in lowly-charged ion collisions (because $2s$ removal is overwhelmed by
additional $2p$ removal for more highly-charged projectiles) one may say that
a fine balance of charge state and structure of a projectile is required to make
ICD a significant process in low-energy capture collisions. 
It would be interesting to see if this can be confirmed by
COBM calculations with effective projectile potentials like the one used in this work
[Eq.~(\ref{eq:vpgsz})] and if 
one can identify an optimal projectile that maximizes
the ICD yield. 
Future work should also be concerned with a more systematic study of the relative
strengths of ICD, CE, and RCT
over a range of projectile species and energies and also for different target systems,
such as water clusters. 
A more quantitative understanding of the ICD process in particular may have
important implications
for ion-beam
cancer therapy, since the low-energy electrons it produces are effective
agents for inflicting cellular damage.

\begin{acknowledgments}
Financial support from the Natural Sciences and Engineering Research Council of Canada (NSERC) 
(RGPIN-2019-06305) is gratefully acknowledged. We thank Amine Cassimi for providing us with 
the experimental and COBM yields listed in Table~\ref{tab1}.
\end{acknowledgments}

\appendix*
\section{}
Let us consider a simplified problem with just two spin-parallel electrons occupying the target and
projectile $2s$ states $\ket{2s^T}$ and $\ket{2s^P}$ before the collision. 
We denote the solutions of the single-particle 
equations for both electrons at a final time $t_f$ after the collision
by $\ket{\psi_{2s^T}}$ and  $\ket{\psi_{2s^P}}$. 
Within the IEM the two-electron problem is represented by a Slater determinant composed of these two
single-particle states.

As shown in Ref.~\cite{hjl85}, when using determinantal wave functions
the inclusive probability for finding one electron in $\ket{2s^T}$
after the collision
while the other one is not observed is given by the one-particle density matrix element
\begin{equation}
	\bra{2s^T}\hat \gamma \ket{2s^T} = 
	|\braket{2s^T}{\psi_{2s^T}}|^2 + |\braket{2s^T}{\psi_{2s^P}}|^2 .
\end{equation}
Not observing one electron implies that it can be anywhere but in the $2s$ target state,
which is blocked by the other electron. Hence, we can interpret
\begin{equation}
	P_{\rm vac}^T \equiv 1 - \bra{2s^T}\hat \gamma \ket{2s^T}
\end{equation}
as the probability for finding the $2s$ target state vacant after the collision.
The principle of detailed balance demands that 
|$\braket{2s^T}{\psi_{2s^P}}|^2 = |\braket{2s^P}{\psi_{2s^T}}|^2$, provided both electrons are
propagated in the same single-particle Hamiltonian. We have checked that our TC-BGM
solutions are consistent with this result.
Accordingly, we can write
\begin{equation}
	P_{\rm vac}^T = 1 - |\braket{2s^T}{\psi_{2s^T}}|^2 - |\braket{2s^P}{\psi_{2s^T}}|^2 ,
\end{equation}
i.e., the $2s$ target vacancy probability is obtained by subtracting the $2s^T\rightarrow 2s^P$ transition
probability from the probability that the initial target electron is not found in its 
initial $2s$ state. 
Given that target excitation is a weak process in the collision system studied in this work, 
we can interpret the latter as the target electron removal probability.

The same argument applies to the initial $2p$ target electrons and can readily be extended
to several target electrons and both spin directions (given that spin flips are impossible 
for a spin-independent Hamiltonian).
This justifies our (approximate) procedure to determine the
`true' single-particle removal probabilities by subtracting
the probabilities for Ne$(2l) \rightarrow $ O$^{3+}(2s)$ from the original Ne$(2l)$ electron removal
probabilities.

%

\bibliography{icd}

\begin{thebibliography}{29}%
\makeatletter
\providecommand \@ifxundefined [1]{%
 \@ifx{#1\undefined}
}%
\providecommand \@ifnum [1]{%
 \ifnum #1\expandafter \@firstoftwo
 \else \expandafter \@secondoftwo
 \fi
}%
\providecommand \@ifx [1]{%
 \ifx #1\expandafter \@firstoftwo
 \else \expandafter \@secondoftwo
 \fi
}%
\providecommand \natexlab [1]{#1}%
\providecommand \enquote  [1]{``#1''}%
\providecommand \bibnamefont  [1]{#1}%
\providecommand \bibfnamefont [1]{#1}%
\providecommand \citenamefont [1]{#1}%
\providecommand \href@noop [0]{\@secondoftwo}%
\providecommand \href [0]{\begingroup \@sanitize@url \@href}%
\providecommand \@href[1]{\@@startlink{#1}\@@href}%
\providecommand \@@href[1]{\endgroup#1\@@endlink}%
\providecommand \@sanitize@url [0]{\catcode `\\12\catcode `\$12\catcode
  `\&12\catcode `\#12\catcode `\^12\catcode `\_12\catcode `\%12\relax}%
\providecommand \@@startlink[1]{}%
\providecommand \@@endlink[0]{}%
\providecommand \url  [0]{\begingroup\@sanitize@url \@url }%
\providecommand \@url [1]{\endgroup\@href {#1}{\urlprefix }}%
\providecommand \urlprefix  [0]{URL }%
\providecommand \Eprint [0]{\href }%
\providecommand \doibase [0]{http://dx.doi.org/}%
\providecommand \selectlanguage [0]{\@gobble}%
\providecommand \bibinfo  [0]{\@secondoftwo}%
\providecommand \bibfield  [0]{\@secondoftwo}%
\providecommand \translation [1]{[#1]}%
\providecommand \BibitemOpen [0]{}%
\providecommand \bibitemStop [0]{}%
\providecommand \bibitemNoStop [0]{.\EOS\space}%
\providecommand \EOS [0]{\spacefactor3000\relax}%
\providecommand \BibitemShut  [1]{\csname bibitem#1\endcsname}%
\let\auto@bib@innerbib\@empty
\bibitem [{\citenamefont {Cederbaum}\ \emph {et~al.}(1997)\citenamefont
  {Cederbaum}, \citenamefont {Zobeley},\ and\ \citenamefont
  {Tarantelli}}]{Cederbaum97}%
  \BibitemOpen
  \bibfield  {author} {\bibinfo {author} {\bibfnamefont {L.~S.}\ \bibnamefont
  {Cederbaum}}, \bibinfo {author} {\bibfnamefont {J.}~\bibnamefont {Zobeley}},
  \ and\ \bibinfo {author} {\bibfnamefont {F.}~\bibnamefont {Tarantelli}},\
  }\href {\doibase 10.1103/PhysRevLett.79.4778} {\bibfield  {journal} {\bibinfo
   {journal} {Phys. Rev. Lett.}\ }\textbf {\bibinfo {volume} {79}},\ \bibinfo
  {pages} {4778} (\bibinfo {year} {1997})}\BibitemShut {NoStop}%
\bibitem [{\citenamefont {Marburger}\ \emph {et~al.}(2003)\citenamefont
  {Marburger}, \citenamefont {Kugeler}, \citenamefont {Hergenhahn},\ and\
  \citenamefont {M\"oller}}]{Marburger03}%
  \BibitemOpen
  \bibfield  {author} {\bibinfo {author} {\bibfnamefont {S.}~\bibnamefont
  {Marburger}}, \bibinfo {author} {\bibfnamefont {O.}~\bibnamefont {Kugeler}},
  \bibinfo {author} {\bibfnamefont {U.}~\bibnamefont {Hergenhahn}}, \ and\
  \bibinfo {author} {\bibfnamefont {T.}~\bibnamefont {M\"oller}},\ }\href
  {\doibase 10.1103/PhysRevLett.90.203401} {\bibfield  {journal} {\bibinfo
  {journal} {Phys. Rev. Lett.}\ }\textbf {\bibinfo {volume} {90}},\ \bibinfo
  {pages} {203401} (\bibinfo {year} {2003})}\BibitemShut {NoStop}%
\bibitem [{\citenamefont {Jahnke}\ \emph {et~al.}(2004)\citenamefont {Jahnke},
  \citenamefont {Czasch}, \citenamefont {Sch\"offler}, \citenamefont
  {Sch\"ossler}, \citenamefont {Knapp}, \citenamefont {K\"asz}, \citenamefont
  {Titze}, \citenamefont {Wimmer}, \citenamefont {Kreidi}, \citenamefont
  {Grisenti}, \citenamefont {Staudte}, \citenamefont {Jagutzki}, \citenamefont
  {Hergenhahn}, \citenamefont {Schmidt-B\"ocking},\ and\ \citenamefont
  {D\"orner}}]{Jahnke04}%
  \BibitemOpen
  \bibfield  {author} {\bibinfo {author} {\bibfnamefont {T.}~\bibnamefont
  {Jahnke}}, \bibinfo {author} {\bibfnamefont {A.}~\bibnamefont {Czasch}},
  \bibinfo {author} {\bibfnamefont {M.~S.}\ \bibnamefont {Sch\"offler}},
  \bibinfo {author} {\bibfnamefont {S.}~\bibnamefont {Sch\"ossler}}, \bibinfo
  {author} {\bibfnamefont {A.}~\bibnamefont {Knapp}}, \bibinfo {author}
  {\bibfnamefont {M.}~\bibnamefont {K\"asz}}, \bibinfo {author} {\bibfnamefont
  {J.}~\bibnamefont {Titze}}, \bibinfo {author} {\bibfnamefont
  {C.}~\bibnamefont {Wimmer}}, \bibinfo {author} {\bibfnamefont
  {K.}~\bibnamefont {Kreidi}}, \bibinfo {author} {\bibfnamefont {R.~E.}\
  \bibnamefont {Grisenti}}, \bibinfo {author} {\bibfnamefont {A.}~\bibnamefont
  {Staudte}}, \bibinfo {author} {\bibfnamefont {O.}~\bibnamefont {Jagutzki}},
  \bibinfo {author} {\bibfnamefont {U.}~\bibnamefont {Hergenhahn}}, \bibinfo
  {author} {\bibfnamefont {H.}~\bibnamefont {Schmidt-B\"ocking}}, \ and\
  \bibinfo {author} {\bibfnamefont {R.}~\bibnamefont {D\"orner}},\ }\href
  {\doibase 10.1103/PhysRevLett.93.163401} {\bibfield  {journal} {\bibinfo
  {journal} {Phys. Rev. Lett.}\ }\textbf {\bibinfo {volume} {93}},\ \bibinfo
  {pages} {163401} (\bibinfo {year} {2004})}\BibitemShut {NoStop}%
\bibitem [{\citenamefont {Jahnke}(2015)}]{Jahnke15}%
  \BibitemOpen
  \bibfield  {author} {\bibinfo {author} {\bibfnamefont {T.}~\bibnamefont
  {Jahnke}},\ }\href {\doibase 10.1088/0953-4075/48/8/082001} {\bibfield
  {journal} {\bibinfo  {journal} {J. Phys. B: At. Mol. Opt. Phys.}\ }\textbf
  {\bibinfo {volume} {48}},\ \bibinfo {pages} {082001} (\bibinfo {year}
  {2015})}\BibitemShut {NoStop}%
\bibitem [{\citenamefont {Ren}\ \emph {et~al.}(2018)\citenamefont {Ren},
  \citenamefont {Wang}, \citenamefont {Skitnevskaya}, \citenamefont {Trofimov},
  \citenamefont {Gokhberg},\ and\ \citenamefont {Dorn}}]{Ren18}%
  \BibitemOpen
  \bibfield  {author} {\bibinfo {author} {\bibfnamefont {X.}~\bibnamefont
  {Ren}}, \bibinfo {author} {\bibfnamefont {E.}~\bibnamefont {Wang}}, \bibinfo
  {author} {\bibfnamefont {A.~D.}\ \bibnamefont {Skitnevskaya}}, \bibinfo
  {author} {\bibfnamefont {A.~B.}\ \bibnamefont {Trofimov}}, \bibinfo {author}
  {\bibfnamefont {K.}~\bibnamefont {Gokhberg}}, \ and\ \bibinfo {author}
  {\bibfnamefont {A.}~\bibnamefont {Dorn}},\ }\href {\doibase
  https://doi.org/10.1038/s41567-018-0214-9} {\bibfield  {journal} {\bibinfo
  {journal} {Nat. Phys.}\ }\textbf {\bibinfo {volume} {14}},\ \bibinfo {pages}
  {1062} (\bibinfo {year} {2018})}\BibitemShut {NoStop}%
\bibitem [{\citenamefont {Iskandar}\ \emph {et~al.}(2015)\citenamefont
  {Iskandar}, \citenamefont {Matsumoto}, \citenamefont {Leredde}, \citenamefont
  {Fl\'echard}, \citenamefont {Gervais}, \citenamefont {Guillous},
  \citenamefont {Hennecart}, \citenamefont {M\'ery}, \citenamefont {Rangama},
  \citenamefont {Zhou}, \citenamefont {Shiromaru},\ and\ \citenamefont
  {Cassimi}}]{Iskandar15}%
  \BibitemOpen
  \bibfield  {author} {\bibinfo {author} {\bibfnamefont {W.}~\bibnamefont
  {Iskandar}}, \bibinfo {author} {\bibfnamefont {J.}~\bibnamefont {Matsumoto}},
  \bibinfo {author} {\bibfnamefont {A.}~\bibnamefont {Leredde}}, \bibinfo
  {author} {\bibfnamefont {X.}~\bibnamefont {Fl\'echard}}, \bibinfo {author}
  {\bibfnamefont {B.}~\bibnamefont {Gervais}}, \bibinfo {author} {\bibfnamefont
  {S.}~\bibnamefont {Guillous}}, \bibinfo {author} {\bibfnamefont
  {D.}~\bibnamefont {Hennecart}}, \bibinfo {author} {\bibfnamefont
  {A.}~\bibnamefont {M\'ery}}, \bibinfo {author} {\bibfnamefont
  {J.}~\bibnamefont {Rangama}}, \bibinfo {author} {\bibfnamefont {C.~L.}\
  \bibnamefont {Zhou}}, \bibinfo {author} {\bibfnamefont {H.}~\bibnamefont
  {Shiromaru}}, \ and\ \bibinfo {author} {\bibfnamefont {A.}~\bibnamefont
  {Cassimi}},\ }\href {\doibase 10.1103/PhysRevLett.114.033201} {\bibfield
  {journal} {\bibinfo  {journal} {Phys. Rev. Lett.}\ }\textbf {\bibinfo
  {volume} {114}},\ \bibinfo {pages} {033201} (\bibinfo {year}
  {2015})}\BibitemShut {NoStop}%
\bibitem [{\citenamefont {Zapukhlyak}\ \emph {et~al.}(2005)\citenamefont
  {Zapukhlyak}, \citenamefont {Kirchner}, \citenamefont {L\"udde},
  \citenamefont {Knoop}, \citenamefont {Morgenstern},\ and\ \citenamefont
  {Hoekstra}}]{tcbgm}%
  \BibitemOpen
  \bibfield  {author} {\bibinfo {author} {\bibfnamefont {M.}~\bibnamefont
  {Zapukhlyak}}, \bibinfo {author} {\bibfnamefont {T.}~\bibnamefont
  {Kirchner}}, \bibinfo {author} {\bibfnamefont {H.~J.}\ \bibnamefont
  {L\"udde}}, \bibinfo {author} {\bibfnamefont {S.}~\bibnamefont {Knoop}},
  \bibinfo {author} {\bibfnamefont {R.}~\bibnamefont {Morgenstern}}, \ and\
  \bibinfo {author} {\bibfnamefont {R.}~\bibnamefont {Hoekstra}},\ }\href
  {\doibase 10.1088/0953-4075/38/14/003} {\bibfield  {journal} {\bibinfo
  {journal} {J. Phys. B}\ }\textbf {\bibinfo {volume} {38}},\ \bibinfo {pages}
  {2353} (\bibinfo {year} {2005})}\BibitemShut {NoStop}%
\bibitem [{\citenamefont {Cybulski}\ and\ \citenamefont
  {Toczyłowski}(1999)}]{Cybulski99}%
  \BibitemOpen
  \bibfield  {author} {\bibinfo {author} {\bibfnamefont {S.~M.}\ \bibnamefont
  {Cybulski}}\ and\ \bibinfo {author} {\bibfnamefont {R.~R.}\ \bibnamefont
  {Toczyłowski}},\ }\href {\doibase 10.1063/1.480430} {\bibfield  {journal}
  {\bibinfo  {journal} {J. Chem. Phys.}\ }\textbf {\bibinfo {volume} {111}},\
  \bibinfo {pages} {10520} (\bibinfo {year} {1999})}\BibitemShut {NoStop}%
\bibitem [{\citenamefont {L\"uhr}\ and\ \citenamefont {Saenz}(2009)}]{Luehr09}%
  \BibitemOpen
  \bibfield  {author} {\bibinfo {author} {\bibfnamefont {A.}~\bibnamefont
  {L\"uhr}}\ and\ \bibinfo {author} {\bibfnamefont {A.}~\bibnamefont {Saenz}},\
  }\href {\doibase 10.1103/PhysRevA.80.022705} {\bibfield  {journal} {\bibinfo
  {journal} {Phys. Rev. A}\ }\textbf {\bibinfo {volume} {80}},\ \bibinfo
  {pages} {022705} (\bibinfo {year} {2009})}\BibitemShut {NoStop}%
\bibitem [{\citenamefont {L\"uhr}\ and\ \citenamefont {Saenz}(2010)}]{Luehr10}%
  \BibitemOpen
  \bibfield  {author} {\bibinfo {author} {\bibfnamefont {A.}~\bibnamefont
  {L\"uhr}}\ and\ \bibinfo {author} {\bibfnamefont {A.}~\bibnamefont {Saenz}},\
  }\href {\doibase 10.1103/PhysRevA.81.010701} {\bibfield  {journal} {\bibinfo
  {journal} {Phys. Rev. A}\ }\textbf {\bibinfo {volume} {81}},\ \bibinfo
  {pages} {010701} (\bibinfo {year} {2010})}\BibitemShut {NoStop}%
\bibitem [{\citenamefont {Engel}\ and\ \citenamefont {Vosko}(1993)}]{ee93}%
  \BibitemOpen
  \bibfield  {author} {\bibinfo {author} {\bibfnamefont {E.}~\bibnamefont
  {Engel}}\ and\ \bibinfo {author} {\bibfnamefont {S.~H.}\ \bibnamefont
  {Vosko}},\ }\href {\doibase 10.1103/PhysRevA.47.2800} {\bibfield  {journal}
  {\bibinfo  {journal} {Phys. Rev. A}\ }\textbf {\bibinfo {volume} {47}},\
  \bibinfo {pages} {2800} (\bibinfo {year} {1993})}\BibitemShut {NoStop}%
\bibitem [{\citenamefont {Engel}\ and\ \citenamefont {Dreizler}(1999)}]{ee99b}%
  \BibitemOpen
  \bibfield  {author} {\bibinfo {author} {\bibfnamefont {E.}~\bibnamefont
  {Engel}}\ and\ \bibinfo {author} {\bibfnamefont {R.~M.}\ \bibnamefont
  {Dreizler}},\ }\href
  {http://resolver.scholarsportal.info/resolve/01928651/v20i0001/31_fetidf}
  {\bibfield  {journal} {\bibinfo  {journal} {J. Comput. Chem.}\ }\textbf
  {\bibinfo {volume} {20}},\ \bibinfo {pages} {31} (\bibinfo {year}
  {1999})}\BibitemShut {NoStop}%
\bibitem [{\citenamefont {Green}\ \emph {et~al.}(1969)\citenamefont {Green},
  \citenamefont {Sellin},\ and\ \citenamefont {Zachor}}]{Green69}%
  \BibitemOpen
  \bibfield  {author} {\bibinfo {author} {\bibfnamefont {A.~E.~S.}\
  \bibnamefont {Green}}, \bibinfo {author} {\bibfnamefont {D.~L.}\ \bibnamefont
  {Sellin}}, \ and\ \bibinfo {author} {\bibfnamefont {A.~S.}\ \bibnamefont
  {Zachor}},\ }\href {\doibase 10.1103/PhysRev.184.1} {\bibfield  {journal}
  {\bibinfo  {journal} {Phys. Rev.}\ }\textbf {\bibinfo {volume} {184}},\
  \bibinfo {pages} {1} (\bibinfo {year} {1969})}\BibitemShut {NoStop}%
\bibitem [{\citenamefont {Szydlik}\ and\ \citenamefont
  {Green}(1974)}]{Szydlik74}%
  \BibitemOpen
  \bibfield  {author} {\bibinfo {author} {\bibfnamefont {P.~P.}\ \bibnamefont
  {Szydlik}}\ and\ \bibinfo {author} {\bibfnamefont {A.~E.~S.}\ \bibnamefont
  {Green}},\ }\href {\doibase 10.1103/PhysRevA.9.1885} {\bibfield  {journal}
  {\bibinfo  {journal} {Phys. Rev. A}\ }\textbf {\bibinfo {volume} {9}},\
  \bibinfo {pages} {1885} (\bibinfo {year} {1974})}\BibitemShut {NoStop}%
\bibitem [{\citenamefont {Kirchner}\ \emph {et~al.}(1998)\citenamefont
  {Kirchner}, \citenamefont {Guly\'as}, \citenamefont {L\"udde}, \citenamefont
  {Engel},\ and\ \citenamefont {Dreizler}}]{tom98}%
  \BibitemOpen
  \bibfield  {author} {\bibinfo {author} {\bibfnamefont {T.}~\bibnamefont
  {Kirchner}}, \bibinfo {author} {\bibfnamefont {L.}~\bibnamefont {Guly\'as}},
  \bibinfo {author} {\bibfnamefont {H.~J.}\ \bibnamefont {L\"udde}}, \bibinfo
  {author} {\bibfnamefont {E.}~\bibnamefont {Engel}}, \ and\ \bibinfo {author}
  {\bibfnamefont {R.~M.}\ \bibnamefont {Dreizler}},\ }\href {\doibase
  10.1103/PhysRevA.58.2063} {\bibfield  {journal} {\bibinfo  {journal} {Phys.
  Rev. A}\ }\textbf {\bibinfo {volume} {58}},\ \bibinfo {pages} {2063}
  (\bibinfo {year} {1998})}\BibitemShut {NoStop}%
\bibitem [{\citenamefont {L\"udde}\ \emph {et~al.}(2018)\citenamefont
  {L\"udde}, \citenamefont {Horbatsch},\ and\ \citenamefont
  {Kirchner}}]{hjl18}%
  \BibitemOpen
  \bibfield  {author} {\bibinfo {author} {\bibfnamefont {H.~J.}\ \bibnamefont
  {L\"udde}}, \bibinfo {author} {\bibfnamefont {M.}~\bibnamefont {Horbatsch}},
  \ and\ \bibinfo {author} {\bibfnamefont {T.}~\bibnamefont {Kirchner}},\
  }\href {https://doi.org/10.1140/epjb/e2018-90165-x} {\bibfield  {journal}
  {\bibinfo  {journal} {Eur. Phys. J. B}\ }\textbf {\bibinfo {volume} {91}},\
  \bibinfo {pages} {99} (\bibinfo {year} {2018})}\BibitemShut {NoStop}%
\bibitem [{\citenamefont {L\"udde}\ and\ \citenamefont
  {Dreizler}(1985)}]{hjl85}%
  \BibitemOpen
  \bibfield  {author} {\bibinfo {author} {\bibfnamefont {H.~J.}\ \bibnamefont
  {L\"udde}}\ and\ \bibinfo {author} {\bibfnamefont {R.~M.}\ \bibnamefont
  {Dreizler}},\ }\href {http://stacks.iop.org/0022-3700/18/i=1/a=012}
  {\bibfield  {journal} {\bibinfo  {journal} {J. Phys. B}\ }\textbf {\bibinfo
  {volume} {18}},\ \bibinfo {pages} {107} (\bibinfo {year} {1985})}\BibitemShut
  {NoStop}%
\bibitem [{\citenamefont {Cassimi}\ \emph {et~al.}(2019)\citenamefont
  {Cassimi}, \citenamefont {Fl\'{e}chard}, \citenamefont {Gervais},
  \citenamefont {M\'{e}ry},\ and\ \citenamefont {Rangama}}]{Cassimi19}%
  \BibitemOpen
  \bibfield  {author} {\bibinfo {author} {\bibfnamefont {A.}~\bibnamefont
  {Cassimi}}, \bibinfo {author} {\bibfnamefont {X.}~\bibnamefont
  {Fl\'{e}chard}}, \bibinfo {author} {\bibfnamefont {B.}~\bibnamefont
  {Gervais}}, \bibinfo {author} {\bibfnamefont {A.}~\bibnamefont {M\'{e}ry}}, \
  and\ \bibinfo {author} {\bibfnamefont {J.}~\bibnamefont {Rangama}},\ }in\
  \href {\doibase https://doi.org/10.1515/9783110580297-009} {\emph {\bibinfo
  {booktitle} {Ion-Atom Collisions: The Few-Body Problem in Dynamic
  Systems}}},\ \bibinfo {editor} {edited by\ \bibinfo {editor} {\bibfnamefont
  {M.}~\bibnamefont {Schulz}}}\ (\bibinfo  {publisher} {De Gruyter, Berlin},\
  \bibinfo {year} {2019})\ p.\ \bibinfo {pages} {213}\BibitemShut {NoStop}%
\bibitem [{\citenamefont {McGuire}(1997)}]{McGuire97}%
  \BibitemOpen
  \bibfield  {author} {\bibinfo {author} {\bibfnamefont {J.~H.}\ \bibnamefont
  {McGuire}},\ }\href@noop {} {\emph {\bibinfo {title} {Electron Correlation
  Dynamics in Atomic Collisions}}}\ (\bibinfo  {publisher} {Cambridge
  University Press, Cambridge},\ \bibinfo {year} {1997})\BibitemShut {NoStop}%
\bibitem [{\citenamefont {Sant'Anna}\ \emph {et~al.}(1998)\citenamefont
  {Sant'Anna}, \citenamefont {Montenegro},\ and\ \citenamefont
  {McGuire}}]{santanna98}%
  \BibitemOpen
  \bibfield  {author} {\bibinfo {author} {\bibfnamefont {M.~M.}\ \bibnamefont
  {Sant'Anna}}, \bibinfo {author} {\bibfnamefont {E.~C.}\ \bibnamefont
  {Montenegro}}, \ and\ \bibinfo {author} {\bibfnamefont {J.~H.}\ \bibnamefont
  {McGuire}},\ }\href {\doibase 10.1103/PhysRevA.58.2148} {\bibfield  {journal}
  {\bibinfo  {journal} {Phys. Rev. A}\ }\textbf {\bibinfo {volume} {58}},\
  \bibinfo {pages} {2148} (\bibinfo {year} {1998})}\BibitemShut {NoStop}%
\bibitem [{\citenamefont {Montanari}\ and\ \citenamefont
  {Miraglia}(2012)}]{montanari12}%
  \BibitemOpen
  \bibfield  {author} {\bibinfo {author} {\bibfnamefont {C.~C.}\ \bibnamefont
  {Montanari}}\ and\ \bibinfo {author} {\bibfnamefont {J.~E.}\ \bibnamefont
  {Miraglia}},\ }\href {\doibase 10.1088/0953-4075/45/10/105201} {\bibfield
  {journal} {\bibinfo  {journal} {J. Phys. B}\ }\textbf {\bibinfo {volume}
  {45}},\ \bibinfo {pages} {105201} (\bibinfo {year} {2012})}\BibitemShut
  {NoStop}%
\bibitem [{\citenamefont {Kirchner}\ \emph {et~al.}(2014)\citenamefont
  {Kirchner}, \citenamefont {Khazai},\ and\ \citenamefont {Guly\'as}}]{tom14}%
  \BibitemOpen
  \bibfield  {author} {\bibinfo {author} {\bibfnamefont {T.}~\bibnamefont
  {Kirchner}}, \bibinfo {author} {\bibfnamefont {N.}~\bibnamefont {Khazai}}, \
  and\ \bibinfo {author} {\bibfnamefont {L.}~\bibnamefont {Guly\'as}},\ }\href
  {\doibase 10.1103/PhysRevA.89.062702} {\bibfield  {journal} {\bibinfo
  {journal} {Phys. Rev. A}\ }\textbf {\bibinfo {volume} {89}},\ \bibinfo
  {pages} {062702} (\bibinfo {year} {2014})}\BibitemShut {NoStop}%
\bibitem [{\citenamefont {Leung}\ and\ \citenamefont
  {Kirchner}(2017)}]{leung17}%
  \BibitemOpen
  \bibfield  {author} {\bibinfo {author} {\bibfnamefont {A.~C.~K.}\
  \bibnamefont {Leung}}\ and\ \bibinfo {author} {\bibfnamefont
  {T.}~\bibnamefont {Kirchner}},\ }\href {\doibase 10.1103/PhysRevA.95.042703}
  {\bibfield  {journal} {\bibinfo  {journal} {Phys. Rev. A}\ }\textbf {\bibinfo
  {volume} {95}},\ \bibinfo {pages} {042703} (\bibinfo {year}
  {2017})}\BibitemShut {NoStop}%
\bibitem [{\citenamefont {Terekhin}\ \emph {et~al.}(2018)\citenamefont
  {Terekhin}, \citenamefont {Quinto}, \citenamefont {Monti}, \citenamefont
  {Foj{\'{o}}n},\ and\ \citenamefont {Rivarola}}]{terekhin18}%
  \BibitemOpen
  \bibfield  {author} {\bibinfo {author} {\bibfnamefont {P.~N.}\ \bibnamefont
  {Terekhin}}, \bibinfo {author} {\bibfnamefont {M.~A.}\ \bibnamefont
  {Quinto}}, \bibinfo {author} {\bibfnamefont {J.~M.}\ \bibnamefont {Monti}},
  \bibinfo {author} {\bibfnamefont {O.~A.}\ \bibnamefont {Foj{\'{o}}n}}, \ and\
  \bibinfo {author} {\bibfnamefont {R.~D.}\ \bibnamefont {Rivarola}},\ }\href
  {\doibase 10.1088/1361-6455/aadb22} {\bibfield  {journal} {\bibinfo
  {journal} {J. Phys. B}\ }\textbf {\bibinfo {volume} {51}},\ \bibinfo {pages}
  {235201} (\bibinfo {year} {2018})}\BibitemShut {NoStop}%
\bibitem [{\citenamefont {Iskandar}\ \emph {et~al.}(2018)\citenamefont
  {Iskandar}, \citenamefont {Fl\'echard}, \citenamefont {Matsumoto},
  \citenamefont {Leredde}, \citenamefont {Guillous}, \citenamefont {Hennecart},
  \citenamefont {Rangama}, \citenamefont {M\'ery}, \citenamefont {Gervais},
  \citenamefont {Shiromaru},\ and\ \citenamefont {Cassimi}}]{Iskandar18}%
  \BibitemOpen
  \bibfield  {author} {\bibinfo {author} {\bibfnamefont {W.}~\bibnamefont
  {Iskandar}}, \bibinfo {author} {\bibfnamefont {X.}~\bibnamefont
  {Fl\'echard}}, \bibinfo {author} {\bibfnamefont {J.}~\bibnamefont
  {Matsumoto}}, \bibinfo {author} {\bibfnamefont {A.}~\bibnamefont {Leredde}},
  \bibinfo {author} {\bibfnamefont {S.}~\bibnamefont {Guillous}}, \bibinfo
  {author} {\bibfnamefont {D.}~\bibnamefont {Hennecart}}, \bibinfo {author}
  {\bibfnamefont {J.}~\bibnamefont {Rangama}}, \bibinfo {author} {\bibfnamefont
  {A.}~\bibnamefont {M\'ery}}, \bibinfo {author} {\bibfnamefont
  {B.}~\bibnamefont {Gervais}}, \bibinfo {author} {\bibfnamefont
  {H.}~\bibnamefont {Shiromaru}}, \ and\ \bibinfo {author} {\bibfnamefont
  {A.}~\bibnamefont {Cassimi}},\ }\href {\doibase 10.1103/PhysRevA.98.012701}
  {\bibfield  {journal} {\bibinfo  {journal} {Phys. Rev. A}\ }\textbf {\bibinfo
  {volume} {98}},\ \bibinfo {pages} {012701} (\bibinfo {year}
  {2018})}\BibitemShut {NoStop}%
\bibitem [{\citenamefont {L\"udde}\ \emph {et~al.}(2016)\citenamefont
  {L\"udde}, \citenamefont {Achenbach}, \citenamefont {Kalkbrenner},
  \citenamefont {Jankowiak},\ and\ \citenamefont {Kirchner}}]{hjl16}%
  \BibitemOpen
  \bibfield  {author} {\bibinfo {author} {\bibfnamefont {H.~J.}\ \bibnamefont
  {L\"udde}}, \bibinfo {author} {\bibfnamefont {A.}~\bibnamefont {Achenbach}},
  \bibinfo {author} {\bibfnamefont {T.}~\bibnamefont {Kalkbrenner}}, \bibinfo
  {author} {\bibfnamefont {H.-C.}\ \bibnamefont {Jankowiak}}, \ and\ \bibinfo
  {author} {\bibfnamefont {T.}~\bibnamefont {Kirchner}},\ }\href
  {https://doi.org/10.1140/epjd/e2016-70097-5} {\bibfield  {journal} {\bibinfo
  {journal} {Eur. Phys. J. D}\ }\textbf {\bibinfo {volume} {70}},\ \bibinfo
  {pages} {82} (\bibinfo {year} {2016})}\BibitemShut {NoStop}%
\bibitem [{\citenamefont {L\"udde}\ \emph {et~al.}(2020)\citenamefont
  {L\"udde}, \citenamefont {Kalkbrenner}, \citenamefont {Horbatsch},\ and\
  \citenamefont {Kirchner}}]{hjl20}%
  \BibitemOpen
  \bibfield  {author} {\bibinfo {author} {\bibfnamefont {H.~J.}\ \bibnamefont
  {L\"udde}}, \bibinfo {author} {\bibfnamefont {T.}~\bibnamefont
  {Kalkbrenner}}, \bibinfo {author} {\bibfnamefont {M.}~\bibnamefont
  {Horbatsch}}, \ and\ \bibinfo {author} {\bibfnamefont {T.}~\bibnamefont
  {Kirchner}},\ }\href {\doibase 10.1103/PhysRevA.101.062709} {\bibfield
  {journal} {\bibinfo  {journal} {Phys. Rev. A}\ }\textbf {\bibinfo {volume}
  {101}},\ \bibinfo {pages} {062709} (\bibinfo {year} {2020})}\BibitemShut
  {NoStop}%
\bibitem [{\citenamefont {Belki\ifmmode~\acute{c}\else \'{c}\fi{}}\ \emph
  {et~al.}(2008)\citenamefont {Belki\ifmmode~\acute{c}\else \'{c}\fi{}},
  \citenamefont {Man\ifmmode~\check{c}\else \v{c}\fi{}ev},\ and\ \citenamefont
  {Hanssen}}]{Belkic08}%
  \BibitemOpen
  \bibfield  {author} {\bibinfo {author} {\bibfnamefont {D.~c.~v.}\
  \bibnamefont {Belki\ifmmode~\acute{c}\else \'{c}\fi{}}}, \bibinfo {author}
  {\bibfnamefont {I.}~\bibnamefont {Man\ifmmode~\check{c}\else \v{c}\fi{}ev}},
  \ and\ \bibinfo {author} {\bibfnamefont {J.}~\bibnamefont {Hanssen}},\ }\href
  {\doibase 10.1103/RevModPhys.80.249} {\bibfield  {journal} {\bibinfo
  {journal} {Rev. Mod. Phys.}\ }\textbf {\bibinfo {volume} {80}},\ \bibinfo
  {pages} {249} (\bibinfo {year} {2008})}\BibitemShut {NoStop}%
\bibitem [{\citenamefont {Aumayr}\ \emph {et~al.}(2019)\citenamefont {Aumayr},
  \citenamefont {Ueda}, \citenamefont {Sokell}, \citenamefont {Schippers},
  \citenamefont {Sadeghpour}, \citenamefont {Merkt}, \citenamefont {Gallagher},
  \citenamefont {Dunning}, \citenamefont {Scheier}, \citenamefont {Echt},
  \citenamefont {Kirchner}, \citenamefont {Fritzsche}, \citenamefont
  {Surzhykov}, \citenamefont {Ma}, \citenamefont {Rivarola}, \citenamefont
  {Fojon}, \citenamefont {Tribedi}, \citenamefont {Lamour}, \citenamefont
  {L{\'{o}}pez-Urrutia}, \citenamefont {Litvinov}, \citenamefont {Shabaev},
  \citenamefont {Cederquist}, \citenamefont {Zettergren}, \citenamefont
  {Schleberger}, \citenamefont {Wilhelm}, \citenamefont {Azuma}, \citenamefont
  {Boduch}, \citenamefont {Schmidt},\ and\ \citenamefont
  {St\"ohlker}}]{Aumayr19}%
  \BibitemOpen
  \bibfield  {author} {\bibinfo {author} {\bibfnamefont {F.}~\bibnamefont
  {Aumayr}}, \bibinfo {author} {\bibfnamefont {K.}~\bibnamefont {Ueda}},
  \bibinfo {author} {\bibfnamefont {E.}~\bibnamefont {Sokell}}, \bibinfo
  {author} {\bibfnamefont {S.}~\bibnamefont {Schippers}}, \bibinfo {author}
  {\bibfnamefont {H.}~\bibnamefont {Sadeghpour}}, \bibinfo {author}
  {\bibfnamefont {F.}~\bibnamefont {Merkt}}, \bibinfo {author} {\bibfnamefont
  {T.~F.}\ \bibnamefont {Gallagher}}, \bibinfo {author} {\bibfnamefont {F.~B.}\
  \bibnamefont {Dunning}}, \bibinfo {author} {\bibfnamefont {P.}~\bibnamefont
  {Scheier}}, \bibinfo {author} {\bibfnamefont {O.}~\bibnamefont {Echt}},
  \bibinfo {author} {\bibfnamefont {T.}~\bibnamefont {Kirchner}}, \bibinfo
  {author} {\bibfnamefont {S.}~\bibnamefont {Fritzsche}}, \bibinfo {author}
  {\bibfnamefont {A.}~\bibnamefont {Surzhykov}}, \bibinfo {author}
  {\bibfnamefont {X.}~\bibnamefont {Ma}}, \bibinfo {author} {\bibfnamefont
  {R.}~\bibnamefont {Rivarola}}, \bibinfo {author} {\bibfnamefont
  {O.}~\bibnamefont {Fojon}}, \bibinfo {author} {\bibfnamefont
  {L.}~\bibnamefont {Tribedi}}, \bibinfo {author} {\bibfnamefont
  {E.}~\bibnamefont {Lamour}}, \bibinfo {author} {\bibfnamefont {J.~R.~C.}\
  \bibnamefont {L{\'{o}}pez-Urrutia}}, \bibinfo {author} {\bibfnamefont
  {Y.~A.}\ \bibnamefont {Litvinov}}, \bibinfo {author} {\bibfnamefont
  {V.}~\bibnamefont {Shabaev}}, \bibinfo {author} {\bibfnamefont
  {H.}~\bibnamefont {Cederquist}}, \bibinfo {author} {\bibfnamefont
  {H.}~\bibnamefont {Zettergren}}, \bibinfo {author} {\bibfnamefont
  {M.}~\bibnamefont {Schleberger}}, \bibinfo {author} {\bibfnamefont {R.~A.}\
  \bibnamefont {Wilhelm}}, \bibinfo {author} {\bibfnamefont {T.}~\bibnamefont
  {Azuma}}, \bibinfo {author} {\bibfnamefont {P.}~\bibnamefont {Boduch}},
  \bibinfo {author} {\bibfnamefont {H.~T.}\ \bibnamefont {Schmidt}}, \ and\
  \bibinfo {author} {\bibfnamefont {T.}~\bibnamefont {St\"ohlker}},\ }\href
  {\doibase 10.1088/1361-6455/ab26ea} {\bibfield  {journal} {\bibinfo
  {journal} {J. Phys. B: At. Mol. Opt. Phys.}\ }\textbf {\bibinfo {volume}
  {52}},\ \bibinfo {pages} {171003} (\bibinfo {year} {2019})}\BibitemShut
  {NoStop}%
\end{thebibliography}%

\end{document}